\documentclass[preprint, times]{aastex62}
\usepackage{CJKutf8}

\usepackage{amsmath}
\usepackage{xspace}
\usepackage{graphicx}
\usepackage{epstopdf}
\usepackage{placeins}

\newcommand{\nq}{$^{14}$N\xspace}
\newcommand{\nc}{$^{15}$N\xspace}

\newcommand{\mn}{N$_2$\xspace}
\newcommand{\ncmn}{$N_{\rm N_2}$\xspace}

\newcommand{\mnq}{$^{14}$\mn\xspace}
\newcommand{\mnc}{N$^{15}$N\xspace}
\newcommand{\nqcr}{$^{14}$N/$^{15}$N\xspace}
\newcommand{\hcnqcr}{HC$^{14}$N/HC$^{15}$N\xspace}
\newcommand{\cnqcr}{C$^{14}$N/C$^{15}$N\xspace}
\newcommand{\enqcr}{[$^{14}$N/$^{15}$N]$_{\rm elem}$\xspace}
\newcommand{\ector}{[C/O]$_{\rm elem}$}

\shorttitle{Nitrogen Fractionation in the TW Hya disk}
\shortauthors{Lee et al.}

\begin{document}

\title{Modeling Nitrogen Fractionation in the Protoplanetary Disk around
TW Hya: Model Constraints on Grain Population and Carbon-to-Oxygen Elemental 
Abundance Ratio}

\author{Seokho Lee}
\affiliation{National Astronomical Observatory of Japan, 2-21-1 Osawa, Mitaka, Tokyo 181-8588, Japan}
\email{seokho.lee@nao.ac.jp}
\author{Hideko Nomura}
\affiliation{National Astronomical Observatory of Japan, 2-21-1 Osawa, Mitaka, Tokyo 181-8588, Japan}
\author{Kenji Furuya}
\affiliation{National Astronomical Observatory of Japan, 2-21-1 Osawa, Mitaka, Tokyo 181-8588, Japan}

\author{Jeong-Eun Lee}
\affiliation{School of Space Research, Kyung Hee University, 1732 Deogyeong-daero, Giheung-gu, Yongin-si, Gyeonggi-do, Korea}

\begin{abstract}
 Observations conducted using the Atacama Large Millimeter/submillimeter Array on the protoplanetary disk around TW Hya show 
the nitrogen fractionation of HCN molecules in \hcnqcr$\sim$120 at a radius of 
$\sim$20~AU.
In this study, we investigated the physical and chemical conditions that control this nitrogen 
fractionation process. To this end, a new disk model was developed, in which the 
isotope-selective photodissociation of \mn and isotope-exchange chemical 
reactions have been incorporated.
 Our model can successfully reproduce the observed HCN column density when the elemental abundances of the gas-phase 
carbon and oxygen are depleted by two orders of magnitude 
relative to those in the interstellar medium and carbon is more abundant than 
oxygen (\ector$>$~1). 
 The isotope-selective photodissociation of \mn is the dominant nitrogen 
fractionation process in our models. 
 The observed \hcnqcr ratio, which increases outwards, 
can also be reproduced by the model by assuming that the small dust grains in the atmosphere of the outer disk are 
depleted more than those in the inner disk. This is consistent with  
 grain evolution models, according to which small dust grains are continuously replenished 
in the inner disk due to fragmentation of the large dust grains that 
radially drift from the outer disk.
\end{abstract}


\section{Introduction}
The isotopic ratio of molecules is a powerful tool for investigating the origin 
of solar system materials and for revealing the possible chemical link between the 
solar system and the interstellar medium (ISM) \citep[see e.g.,][for a recent 
review]{Ceccarelli2014}.
Nitrogen has two stable isotopes, \nq and \nc. The nitrogen isotope ratio, 
\nqcr, is nonuniform in the primitive materials of the solar system.
Cometary ices, such as HCN and NH$_3$, have a \nqcr ratio of $\sim$140 
\citep[e.g.,][]{Mumma2011,Shinnaka2016}, which is a factor of three lower 
than the elemental nitrogen isotope ratio, \enqcr, of 440 observed in solar
 winds \citep{Marty2011}; that is, cometary ices are enriched in \nc.
By contrast, bulk meteorites show an isotope ratio between the two aforenoted
values \citep[$\sim$270,][]{Furi2015}. Thus, the origin of nitrogen isotope 
fractionation, i.e., the mechanism and location of fractionation initiation, remains unclear.

 The \nqcr ratios of various molecules have been quantified in the 
low-mass star-forming regions at different evolutionary stages, from the cold 
dense cores to the protoplanetary disks. The \enqcr ratio in the ISM is $\sim$300 
\citep{Lucas1998,Ritchey2015,Hily-Blant2017}. In prestellar cores, nitriles 
such as HCN, HC$_3$N, and HC$_5$N have a \nqcr ratio of $\sim$250 -- 400, 
which is close to \enqcr when the ratios are derived using the direct method rather than 
the double-isotope method \citep[e.g., ][]{Taniguchi2017,Hily-Blant2018,
Magalhaes2018}.
On the other hand, the observed \nqcr ratios in N$_2$H$^+$ and N$_2$D$^+$ in 
the prestellar cores are higher ($\sim$500 -- 1000) than \enqcr 
\citep[][]{Bizzocchi2013,Furuya2018b,Redaelli2018}.
In more evolved sources,  Class 0/I protostellar sources and Class II 
protoplanetary disks, the \nqcr ratio in HCN has been observed 
\citep[e.g.,][]{Guzman2017,Bergner2020}.
The \nqcr ratio in HCN is lower than \enqcr and decreases from 290 to 90 with the 
evolutionary stages traced by bolometric temperature \citep{Bergner2020}.

Two chemical process have been proposed to explain the observed \nc 
fractionation: isotope-exchange reactions \citep{Terzieva2000} and 
isotope-selective photodissociation of \mn \citep{Liang2007,Heays2014}. 
\nc fractionation by isotope-exchange reactions has been thoroughly studied by 
many researchers \citep[e.g.,][]{Terzieva2000,Rodgers2008,Roueff2015}.
A \nc-bearing molecule has a slightly lower zero-point energy than the 
corresponding \nq isotopologue.
This results in endothermicity for the exchange of \nc for \nq, leading to the 
enrichment of \nc in molecules at low temperatures.
However, astrochemical models have found that under the molecular cloud/core 
conditions (i.e., at temperatures as low as 10 K), the isotope-exchange reactions of nitrogen are 
inefficient \citep{Roueff2015,Wirstrom2018,Loison2019} and the molecular \nqcr 
ratios simply reflect the \enqcr values when chemistry reaches the steady state \citep[][]{Roueff2015}.

  Astrochemical models suggest that isotope-selective 
photodissociation of \mn is the more dominant process as compared to the isotope-exchange 
reactions in molecular clouds and protoplanetary disks 
\citep{Furuya2018,Visser2018}.
Around \mn photodissociation front, \mnc is photodissociated more than \mnq, 
which results in the $^{15}$N enrichment of atomic nitrogen, and thus also the other 
N-bearing molecules produced from atomic nitrogen. 
The isotope-selective photodissociation of N$_2$ can explain, at least 
qualitatively, the depletion of \nq in N$_2$H$^+$ in prestellar cores 
\citep{Furuya2018} and the enrichment of \nq in HCN in Class II disks 
\citep{Visser2018}. 
As pointed out in the paper by \citet{Bergner2020}, the decrease in the \nqcr ratio in HCN with the 
evolutionary stages could be naturally explained by the isotope-selective 
photodissociation of N$_2$. This is because the ultraviolet (UV) exposure from the central stars increases 
with the dissipation of the envelope materials.

 The observations, performed using the Atacama Large Millimeter/submillimeter Array (ALMA), of the TW Hya disk provide the spatial variation 
of nitrogen fractionation in gas-phase HCN \citep{Hily-Blant2019}. The 
\nqcr ratio in the comet-forming region ($\sim$20~AU) is 121 $\pm$ 11, which is 
 a factor of three lower than the \enqcr ratio \citep{Hily-Blant2017}. 
This feature is similar to that recorded in  comets, implying that HCN 
could trace the fractionated reservoir observed in  comets. However, whether the observed gas-phase isotope ratio reflects the 
isotope ratio in the ice phase or not
remains unclear.
The observed isotope ratio increases in the outward direction of the disk, which might indicate that the 
isotope-selective photodissociation of \mn is a dominant chemical process, as 
reported based on the disk model by \citet{Visser2018}. However, the disk model 
is unable to reproduce the spatial distribution of nitrogen fractionation
 because in that study, typical disk models were used rather than a dedicated disk model for the
TW Hya disk, even though the radial distribution of nitrogen fractionation strongly 
depends on the physical parameters of the model \citep{Visser2018}.
 Furthermore, they used only the initial abundance, similar to the ISM value, whereas the gas-phase carbon and oxygen are depleted in the TW Hya 
disk \citep[e.g., ][]{Hogerheijde2011,Bergin2013,Favre2013,
Du2015,Kama2016}, which could also affect the nitrogen fractionation.

 In this study, we have developed a new thermochemical disk model incorporating the nitrogen
 isotope chemistry and have attempted to reproduce the nitrogen fractionation of the gas-phase HCN 
observed in the TW Hya disk. In Section~\ref{sec:model}, we describe the physical and 
chemical models for the TW Hya disk. The abundance and isotope profiles obtained using our 
models have been presented in Section~\ref{sec:results}. Our models have been compared with experimental 
observations, and the interpretation of the best-fit model and the effects of 
a few parameters have been discussed in Section~\ref{sec:discussions}. A summary of the 
conclusions drawn from this study has been given in Section~\ref{sec:summary}. 

\section{Model} \label{sec:model}
In this study, we have investigated the nitrogen fractionation observed in the TW Hya disk 
 using a newly developed axisymmetric two-dimensional (2D) thermochemical disk model, packages of
 unified modeling for radiative transfer, gas energetics, and chemistry 
(PURE-C).
 This model has been updated from the thermochemical model for the UV-heated outflow cavity 
walls \citep{Lee2014,Lee2015}. A short summary of the model is provided here and a detailed description is given in Appendix A.

 PURE-C calculates the radiation field and the dust temperature using the 
Monte Carlo method for a given density profile. The gas thermal balance and 
chemistry are solved iteratively.
 In this stage, a simple chemical network is used since only specific species
 such as H, H$_2$, CO, H$_2$O, electrons, and some metal species affect the gas 
thermal balance. The chemistry is evolved with the full chemical network 
for 10 Myrs using the obtained density, temperature, and radiation field 
profiles.

\subsection{Parameters for the TW Hya disk}
\label{subsec:twhya_model}
 The parameters of the TW Hya disk were adopted from the literature. The important parameters along with the symbols 
used in this paper and their adopted values are listed in Table~\ref{tab:table1}.
The density profiles of the gas and dust grains were adopted from \citet{Cleeves2015},
 which fits the spectral energy distribution of the TW Hya disk. In their model, the small 
dust grains having radii $r_g$= 0.005~$\mu$m -- 1~$\mu$m were coupled with gas and the 
large dust grains ($r_g$= 0.005~$\mu$m -- 1~mm) were concentrated near the midplane.
 The gas density profile of the disk ($\rho_g$) was described in the following 
parameterized form in cylindrical coordinates ($R$, $z$) \citep{Hartman1998,
Du2014,Woitke2016}:
\begin{equation}
\rho_g(R, z) =\frac{\Sigma_g(R)}{\sqrt{2 \pi}h(R)}\exp\left[-\frac{1}{2}\left(
	\frac{z}{h(R)}\right)^2\right],
\end{equation}
where the surface density, $\Sigma_g(R)$, and the scale height, $h(R)$, are given by
\begin{eqnarray}
\Sigma_g(R) &=& \Sigma_c\left(\frac{R}{R_c}\right)^{-\gamma} \exp\left[-\left(
	\frac{R}{R_{\rm c}}\right)\right],\\
h(R)&=& h_c \left(\frac{R}{R_c}\right)^\beta.
\end{eqnarray}
The gas density profile of our model is shown in the left panel of 
Figure~\ref{fig:model}. 

 The densities of the small ($\rho_s$) and large ($\rho_l$) dust grain populations 
 were described by
 \begin{equation}
 \rho_s (R,z) = \frac{(1-f) \Sigma_d(R)}{\sqrt{2 \pi}h(R)} \exp{\left[-\frac{1}{2} \left(\frac{z}{h(R)}  \right)^{2} \right]}, 
 \end{equation} and
 \begin{equation}
 \rho_l (R,z) = \frac{f\Sigma_d(R)}{\sqrt{2 \pi} \chi_d h(R)}  \exp{ \left[-\frac{1}{2} \left(\frac{z}{\chi_d h(R)}  \right)^{2} \right]},
 \end{equation}
respectively, where $f$ is the mass fraction of the large dust grains and $\chi_d$ is
 the concentration factor of the large dust grains. 
We assumed a radially constant gas-to-dust ratio of $\delta_g= 100$, which was 
defined by comparing the surface density between the gas and dust grains 
($\delta_g= \Sigma_g/\Sigma_d$).

In this study, we ran three models with different mass fractions of the large 
dust grain. According to \citet{Visser2018}, the nitrogen fractionation of 
HCN is affected predominantly by the grain population among the disk parameters 
 because the grain population affects the UV flux for a given gas 
column density and changes the importance of the isotope-selective 
photodissociation of \mn, which is the dominant process for nitrogen 
fractionation. For the TW Hya disk, $f=0.9$ was used by \citet{Cleeves2015} and  
$f=0.99$ was used by \citet{Kama2016}, where the parameters in the density 
profile were also different from those used in \citet{Cleeves2015}. Both models were able to
 fit the spectral energy distribution of the TW Hya disk.
The model developed by \citet{Cleeves2015} is unable to account for the observed 
radial concentration of large dust grains because the gas-phase species depend 
mainly on the total surface area of the small dust grains in the atmosphere, which are responsible for the  
observed molecular emission lines.
Therefore, we adopted the simplest model of \citet{Cleeves2015} and ran the 
models with two $f$ values of 0.9 (ML) and 0.99 (MS). We also introduced a 
hybrid model (MH) in order to satisfactorily fit the observations (see 
Sect.~\ref{subsec:discuss1});
 \begin{equation}
  \label{eq:frac}
  f= \left\{\begin{array}{cl}
	  0.9,                                 & \;if\,\, R\leq 20 {\rm~AU} \\
	  0.9 + 0.09 \times (R -20 {\rm~AU})/20 {\rm~AU},   & \; if\,\,  20 {\rm AU} \leq R \leq 40 {\rm~AU} \\
	  0.99,                                &  \; if\,\, R > 40 {\rm~AU}.
  \end{array}\right.
 \end{equation}
It should be noted here that the three models exhibit similar spectral energy distributions, except for
 the flux around 200~$\mu$m with the maximum difference of a factor of two.

\subsection{Chemistry for the nitrogen fractionation of HCN in the TW Hya disk} \label{subsec:chemistry}
  Certain features experimentally observed in the disk were taken into account in PURE-C. 
In the midplane of the disk, chemical reactions on the dust grains are important  
for forming complex organic molecules \citep[e.g.,][]{Furuya2014,Walsh2014}. 
 The line-shielding effect of molecules by H, H$_2$, C, and CO was also considered in the model by using 
the tables provided by \citet{Heays2017}, because UV photons can be attenuated
 by abundant molecules as well as the dust grains present in the atmosphere of the disk.
The column densities used in self-shielding and line-shielding were calculated 
in 2D spaces by averaging the radial and vertical column densities weighted 
by the UV fluxes along each direction. 
X-ray ionization is another important process occurring in the disk \citep{Cleeves2015,Offner2019},  
and thus we adopted an analytical model to calculate the X-ray ionization rate in the 
disk \citep{Igea1999}. 
 We assumed that X-ray ionization works similar to cosmic rays in order to incorporate its chemistry 
\citep[see ][]{Walsh2012}, and the cosmic ray 
ionization rate was taken to be 1.0$\times10^{-19}$~s$^{-1}$ \citep{Cleeves2015}.

The nitrogen isotope chemistry was also included in PURE-C.
We reduced the chemical network used in \citet{Furuya2018}, which is modeled for  
nitrogen fractionation in the molecular cloud. The original network 
included the spin state of H$_2$, the isotope-selective photodissociation of \mn 
\citep{Heays2014}, and the isotope-exchange reactions \citep{Roueff2015}. However, 
this network is too large to solve the 2D disk model. Thus, we have focused only on 
the nitrogen fractionation of HCN in this work. \citet{Visser2018} 
used the 2D disk model for nitrogen isotope chemistry with a 
small network including species listed in Table~B.1 of their paper, which were sufficient for 
investigating the nitrogen fractionation of HCN. Therefore, we reduced the 
chemical network of \citet{Furuya2018} by choosing the species used in 
\citet{Visser2018} except for the polycyclic aromatic hydrocarbon (PAH) species and $^{13}$C-bearing species. 
It should be noted here that PURE-C adopts a two-phase model (gas and grain surface) 
using the binding energy on water ice, whereas the model of 
\citet{Furuya2018} adopted the three-phase model (gas, grain surface,
 and ice mantle) using the binding energy corrected by the composition of the ice 
 surface. In our model, the binding energies (on water ice) of \mn, HCN, CN, and 
 NH$_3$ were 1000~K, 3700~K, 2800~K, and 5500~K respectively.

The stellar spectra of the TW Hya disk \citep[e.g.,][]{Nomura2005,France2014} were used for obtaining the unshielded 
photodissociation/ionization rate ($\alpha_0$) and the line-shielding function 
\citep{Heays2017}. The dust shielding function, which is generally calculated 
for an infinite slab, is inadequate for our 2D disk model that considers the 
settling of dust grains. We calculated the dust-attenuated UV fluxes ($\chi$) at one 
representative wavelength of 9.8~eV by solving the UV radiative transfer, and the 
photodissociation/ionization rate was simply calculated as $\alpha_0 \chi$. 
For the CO \citep{Visser2009} and \mn \citep{Heays2014} self-shielding 
functions, we chose the excitation temperatures and Doppler widths of 20 K and 
0.3~km~s$^{-1}$ for CO and 30 K and 0.13~km~s$^{-1}$ for \mn respectively.

In this study, we have investigated the effect of the initial elemental abundances of 
carbon and oxygen.
Elemental gas-phase carbon and oxygen deficiencies of a factor of 10--100 and
their abundance ratios, \ector$>$ 1, were already inferred for the TW Hya disk 
\citep{Hogerheijde2011,Bergin2013,Favre2013,Du2015,Kama2016}. These 
deficiencies could be understood as a chemical and grain evolution
\citep{Kama2016,Bergin2016}: CO and H$_2$O freeze-out onto the dust grains and 
the grains covered with ices settle down towards the midplane and drift inwards.
Chemical reactions also lead to the depletion of the gas-phase CO 
\citep[e.g.,][]{Walsh2010,Furuya2014,Schwarz2018,Schwarz2019}.
 We adopted \enqcr = 330 for the TW~Hya disk 
\citep{Hily-Blant2017}. The \ector\, ratio affects the HCN column density 
\citep{Cleeves2018} and could also possibly affect the nitrogen fractionation 
of HCN.
 There is no study available on the elemental gas-phase nitrogen abundance or 
the [N/O]$_{\rm elem}$ ratio in the TW Hya disk, and the gas-phase nitrogen 
element does not deplete in the IM Lup protoplanetary disk \citep{Cleeves2018}.
Therefore, we modified the initial abundance values reported by \citet{Cleeves2015} and ran  
four models with different initial abundances listed in 
Table~\ref{tab:initabun}. The I1 model represents the typical values in the ISM, 
and the I2 model reduces the abundances of water ice and CO gas by two orders 
of magnitude. In the I3 and I4 models, the initial abundances of 
water ice and atomic carbon were changed so that the elemental abundances
 in the gas phase satisfy \ector$>$~1 after the desorption of water ice.
Therefore, in total, we ran 12 models listed in Table~\ref{tab:modellist} 
using different grain populations (ML, MS, MH) and different initial abundances 
(I1--I4).

\section{Results} \label{sec:results}
\subsection{Physical characteristics}

\begin{figure*}
\includegraphics[width=0.5 \textwidth]{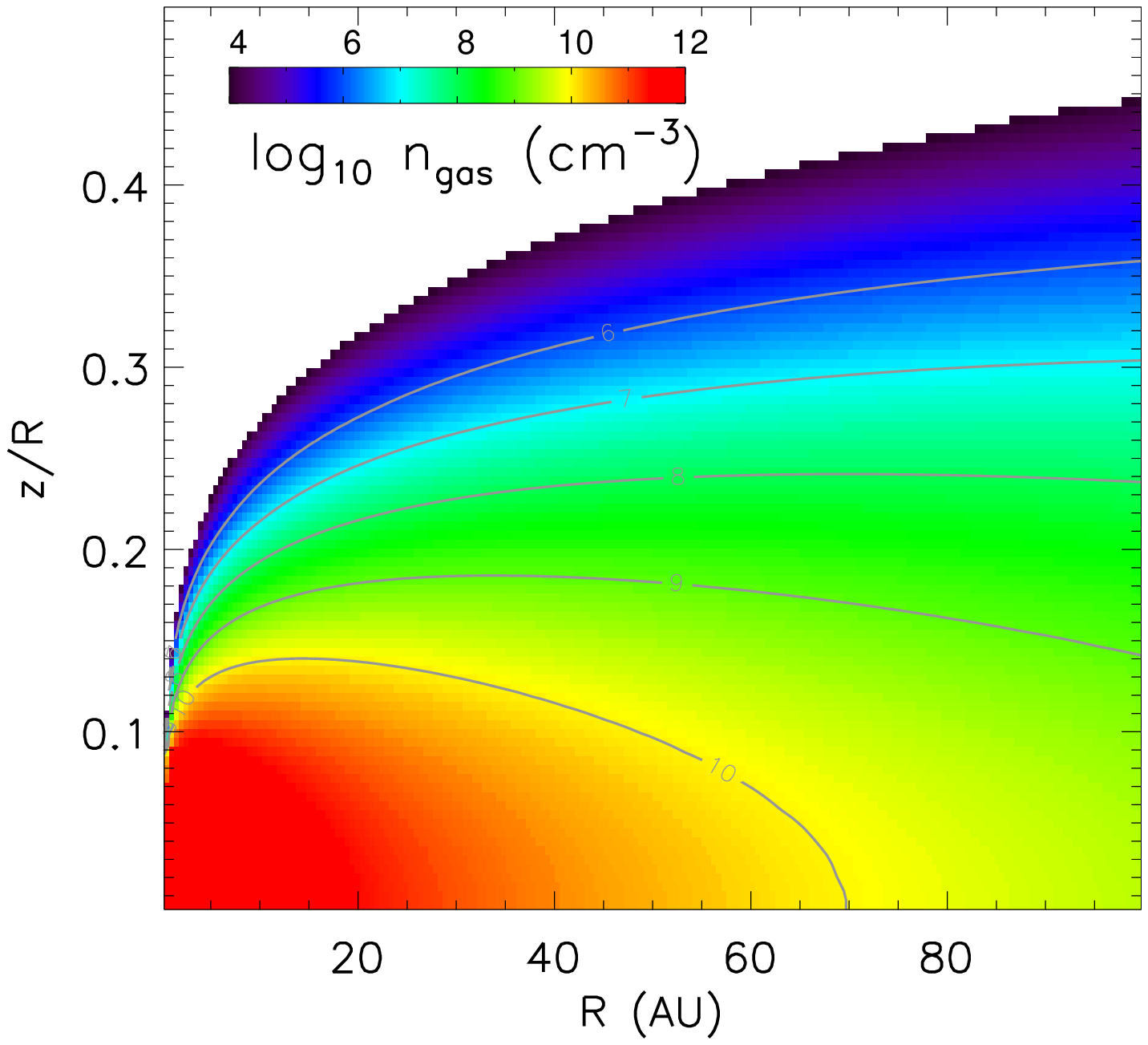}
\includegraphics[width=0.5 \textwidth]{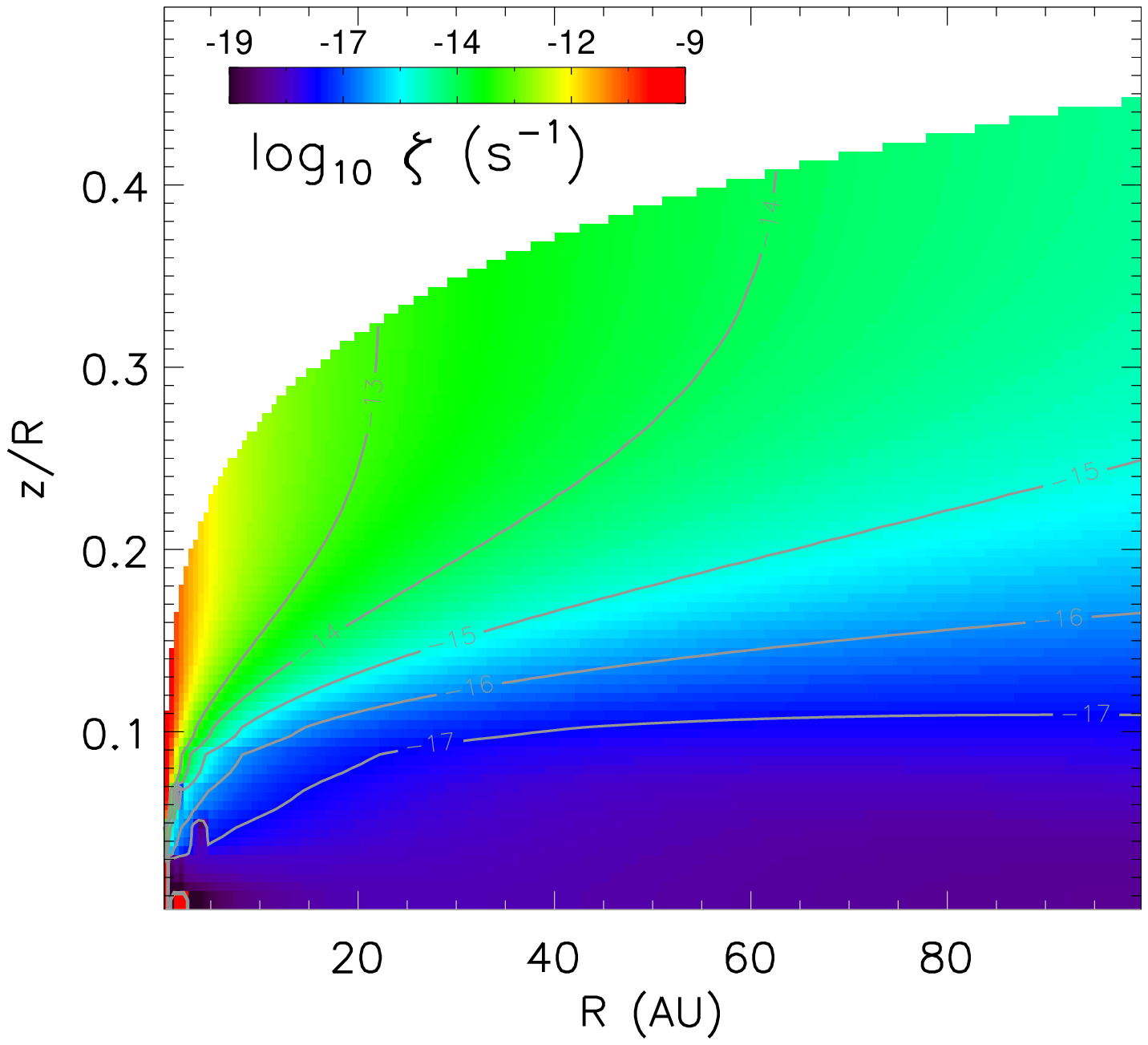}
\caption{Gas density and ionization rate (sum of X-ray and cosmic ray 
	contribution) in all models (see 
Section~\ref{subsec:twhya_model}).
}
\label{fig:model}
\end{figure*}

Figure~\ref{fig:model} shows the 2D distributions of the gas density 
($n_{\rm gas}$) and the ionization rate ($\zeta$), and Figure~\ref{fig:model2} 
presents the profiles of the UV flux ($\chi$, in the units of Draine field), dust 
temperature, gas temperature, and \mn column density from the central star to 
a given position (hereafter \ncmn)\footnote{ In this work, two types of column 
densities have been used. One (\ncmn) is measured from the central star to a given 
position and used for calculating the self- and line-shielding effects (see 
Figures~\ref{fig:model2} and \ref{fig:model_v}, and Appendix). 
The other is integrated vertically to compare with observations, and is
the default column density used in this study (see Figures~\ref{fig:nc} and 
\ref{fig:nc_check}).} 
for the MLI3, MSI3, and MHI3 models. 
Figure~\ref{fig:model_v} shows the vertical cuts of Figures~\ref{fig:model} 
and \ref{fig:model2} for the MLI3 model at three radii of 20, 40, and 
60~AU. The column density ratio of \hcnqcr observed in the TW Hya disk 
\citep{Hily-Blant2019} is the lowest at 20~AU and peaks around 40~AU. 
The 60~AU point is just outside the radius where the ratio is measured (55~AU). 

Grain populations affect the UV fluxes and the dust and gas temperatures. 
Small dust grains in the atmosphere in the MS model are observed to be depleted by an 
order of magnitude compared to those in the ML model. Therefore, UV photons are 
attenuated less in the MS model than in the ML model. Photons in the other 
wavelengths from the central star are also observed to be attenuated less in the MS model,
and thus a higher radiation flux in the MS model produces a higher dust 
temperature than those in the ML model. At a height of $z/R<$~0.2,  
collision between the gas and dust grains is a dominant heating and cooling source.
 Thus, the gas temperature also exhibits a trend similar to the dust temperature.

The grain populations also influence \ncmn as shown in the bottom panels in
 Figure~\ref{fig:model2}. \ncmn depends on the path of the UV photons as well 
as the \mn number density (abundance) distribution. The white dotted lines in the 
bottom panels in Figure~\ref{fig:model2} indicate the height at which the UV fluxes 
directly irradiated from the central star are the same as those 
scattered by the dust grains in the atmosphere of the disk and descending
vertically.
Most UV photons pass via the radial direction from the central star above 
the white dotted lines. Although the \mn number density distribution is also affected by 
\ncmn, the latter tends to be larger for UV photons passing through the radial direction than for those passing via the vertical direction in our models.
The height of the white dotted line in the MS model is lower than that in the
ML model because the small dust grains are less abundant in the former as compared to 
the latter. This causes less scattering of the UV photons in the atmosphere in 
the MS model than in the ML model. 
Therefore, for a given radius and height, the MS model has a higher \ncmn than the 
ML model, and near the midplane in the outer disk, the MS model has a lower \mn 
photodissociation rate than the ML model even though the UV flux is higher in the 
former than in the latter. 
 For example, at a radius of 60~AU and a height of $z/R$=~0.2, the 
MSI3 model has a UV flux and $N_{\rm N_2}$ that are a factor of five higher than the
 MLI3 model. As a result, the \mn photodissociation rate in the MSI3 model is observed to be a 
factor of five lower than that in the MLI3 model, indicating that the \mn 
self-shielding effect is dominant than the dust attenuation of UV photons 
(see the dotted and solid lines in the top-right panel of 
Figure~\ref{fig:model_v}).

\subsection{Cyanide abundances and isotope ratios}
 Figures~\ref{fig:abun} and \ref{fig:isotope} show the distributions of 
the abundance and nitrogen isotope ratio respectively for atomic~N, \mn, HCN, and CN 
in the MLI3 model and the corresponding vertical cuts at 20, 40, and 
60~AU are plotted in Figure~\ref{fig:abun_v}. 
Figure~\ref{fig:abun_v_s} plots the same quantities as Figure~\ref{fig:abun_v} 
except for the MSI3 model.
The gas-phase \mn is observed to be abundant in the warm molecular layer ($z/R\geq 0.1$ and 
UV flux $\chi\geq 1$), and both CN- and HCN-abundant layers are observed to be located within 
the \mn abundant layer.
An upper boundary of the \mn abundant layer is confined by the \mn 
photodissociation front. As for the lower boundary, all N-bearing species 
freeze-out on the dust grains.
 The evaporation temperature of \mn is around 20~K in our models, and the
freeze-out onto the dust grains depletes the gas-phase \mn near the 
midplane in the outer disk ($>$20~AU). The gas-phase \mn is also depleted 
above the \mn snowline with the dust temperature of $\sim$20~K, as shown in the 
top-right plot in Figure~\ref{fig:abun}. In this region, the ionization rate is 
higher than $\sim$10$^{-17}$~s$^{-1}$ (see the right panel of Figure~\ref{fig:model} 
and the bottom-left panel of Figure~\ref{fig:model_v}), and the gas-phase CO is 
converted to CO$_2$ ice within a few Myrs \citep{Furuya2014, Bergin2014}.
The gas-phase \mn becomes NH$_3$ ice after the gas-phase 
CO depletes \citep[see ][]{Furuya2014}. 
The lower boundary of the \mn-abundant layer moves from the dense region closer to
the midplane to the less dense region in the upper layer with the passage of time.
 Notably, in our models with \ector$> 1$, even in the warm molecular 
 layer, HCN ice is more
 abundant than the gas-phase HCN and it is the most dominant ice among the
N-bearing species.

Heavy atomic nitrogen ($^{15}$N) is enriched near the atomic~N and \mn 
transition layer (hereafter, NT layer), where the atomic~N and \mn 
abundances with respect to the total hydrogen nuclei 
($n_{\rm gas}$ = $n$(H) + 2 $n$(H$_2$)) are higher than $\sim$10$^{-7}$ due to the isotope-selective photodissociation 
of \mn. 
Formation of CN and HCN generally begins with atomic~N, and thus the 
nitrogen isotope ratios of the two molecules follow the atomic \nqcr ratio 
near the NT layer. These results are similar to those reported by \citet{Visser2018}. 

The HCN-abundant layer could be composed of two layers: 
(i) an NT layer where the \hcnqcr ratio follows the atomic \nqcr ratio and 
(ii) a ``lower molecular layer" (hereafter LM layer) just below the NT 
layer with an abundance of the atomic~N lower than $\sim$10$^{-7}$. 
 In this layer, the isotope-exchange reactions affect the \hcnqcr ratio. In our model, we included
 the reaction of $^{15}$N + CN $\leftrightarrow$ $^{14}$N + C$^{15}$N, which
 was ignored in the work of \citet{Visser2018}. 
For the rate coefficient of this reaction, we used the upper limit value 
proposed by \citet{Roueff2015}. As can be seen in the bottom panels of 
Figure~\ref{fig:abun_v}, the atomic \nqcr ratio is higher than \enqcr, whereas
 the \hcnqcr ratio is lower than \enqcr in the LM layer.
This is due to the abovementioned isotope-exchange reactions. 
However, the \hcnqcr ratio in the LM layer is closer to the \enqcr ratio 
compared to that in the NT layer and the isotope-selective photodissociation of
 \mn is the dominant process for nitrogen fractionation of HCN. 
The CN-abundant layer also exhibits a trend similar to the HCN-abundant layer.
 Furthermore, the CN-abundant layer has an additional ``upper 
molecular layer" just above the NT layer, where the \cnqcr ratio is close to 
the \enqcr ratio because the \ncmn is too low, and thus 
the isotope-selective photodissociation of \mn does not work.

\subsection{Effect of the initial abundance and the grain population}
The column densities of HCN, CN, and C$_2$H are affected more by the \ector\, 
ratio than by the elemental abundances of carbon and oxygen and the dust grain 
population. Figure~\ref{fig:nc} shows the column densities of HCN, CN, and 
C$_2$H and the column density ratios of \hcnqcr and \cnqcr obtained from our models with 
different grain populations and initial abundances (gas-phase elemental 
abundances and the \ector\, ratio). 
 The column density of C$_2$H is  more sensitive to the \ector\, ratio 
 than those of HCN and CN \citep{Cleeves2018}. Further, the models with 
\ector$>$1 can reproduce the observationally detectable column density of 
C$_2$H, as reported by \citet{Bergin2016}. 
 The bright C$_2$H emission in the protoplanetary disk is used as an 
indicator of \ector$>$1, and thus we also checked the effect of the initial 
abundance and grain population on the C$_2$H column density and compared 
them with observations in this 
study (see Section~\ref{subsec:discuss1}).
 The HCN and CN column densities for the models with \ector$>$~1 are a 
factor of 10 -- 100 higher than those for the models with \ector$<$~1 
\citep{Cleeves2018}.  
 For the models with \ector$>$~1, the MS models show a factor of 
2 -- 10 higher column densities of HCN, CN, and C$_2$H than the ML models. The 
UV photons penetrate closer to the midplane of the disk in the MS model than in the 
ML model and this affects the gas-phase CO and \mn depletion processes.
When the models have the same \ector\, ratio, the gas-phase elemental 
abundances of carbon and oxygen do not affect the HCN and CN column densities, 
whereas the C$_2$H column density obtained from the MLI4 model is an order of magnitude 
higher than that obtained from the MLI3 model (see the green and red lines in 
Figure~\ref{fig:nc}).

On the other hand, the column density ratios of \hcnqcr and \cnqcr are affected
 by the grain population as well as the \ector\, ratio.  
 The column density ratio of the isotopologues is determined by the 
contribution of the NT layer relative to that of the LM layer. 
The NT layer reduces the column density ratio of the isotopologues, whereas the LM 
layer leads to the ratio to be close to the \enqcr ratio. 
 For HCN, the models with \ector$>$~1 have higher ratios than the models 
with \ector$<$~1. 
In the models with \ector$<$~1, more atomic O exists in the 
gas phase and atomic~N becomes NO rather than HCN and CN. This effect is found to be greater in the LM layer than in the NT layer, and thus the contribution of the 
NT layer to the total column density in the model with \ector$>$~1 is  
 smaller than in the model with \ector$<$~1. 
 The contribution of the ``upper molecular" layer to the total CN column 
density is small in our models except for the models having an initial abundance
 of I1. Therefore, the I1 models have a column density ratio of \cnqcr that is higher
 or comparable to that in the models with \ector$>$~1.

 The column density ratios of \hcnqcr and \cnqcr obtained from the MS model are 
 closer to \enqcr than those obtained from the ML model. As mentioned above, the 
UV photons penetrate closer to the midplane of the disk in the MS model as compared to that in the 
ML model, and the LM layer is extended to a lower height in the MS model 
than in the ML model (see Figures \ref{fig:abun_v} and \ref{fig:abun_v_s}).
 Furthermore, \ncmn in the NT layer increases faster from the disk 
atmosphere to the deeper regions in the MS model than in the ML model, as shown
 in the bottom panels in Figure~\ref{fig:model2}, because the UV photons in the NT 
layer tend to propagate radially in the MS model, whereas they propagate vertically in the ML model
 (see the dotted vertical lines in Figures~\ref{fig:abun_v} and 
 \ref{fig:abun_v_s}).
 Therefore, the isotope-selective photodissociation of \mn works in the 
narrower height ranges in the MS model than in the ML model. This indicates that 
the width of the NT layer in the MS model is narrower than that in the ML model 
 (see Figures \ref{fig:abun_v} and \ref{fig:abun_v_s}).
 As a result, the contribution of the NT layer in the MS model is smaller than that
in the ML model, and thus the isotope ratios are closer to \enqcr in the 
MS model than in the ML model (see Figure~\ref{fig:nc}).  
 It should be noted that this trend is contrary to the result obtained by \citet{Visser2018}
where the model having a larger fraction of mass in the large dust grains has a lower 
column density ratio of \hcnqcr than the model having a smaller fraction of 
mass in the large dust grains (see Fig. 14 in \citet{Visser2018}). 
 The main difference between our model and theirs is the method of 
calculating the column density under the self-shielding effect.
 In their model, the column density is measured as the minimum of the radial/vertical 
 (inward/upward) column density \citep[see also ][]{Miotello2014}. 
 On the other hand, the column density weighted by the UV flux in the radial and 
vertical directions has been used in our model (see Appendix for more details). 
Therefore, the value of \ncmn obtained using our model tends to be higher than that obtained in their work and depends on 
the path of the UV photons, which could result in opposite trends in the 
two models.

\section{Discussion} \label{sec:discussions}
\subsection{Comparison with observations} \label{subsec:discuss1}
 A comparison of the results obtained from our models has been done with the observations on the TW Hya disk.
The observed column densities derived from the single dish observations 
\citep[gray bars, ][]{Kastner2014} and the ALMA observations (see below) 
are plotted in Figure~\ref{fig:nc}. 
The observed isotope column density ratios of HCN and CN were adopted from 
\citep{Hily-Blant2017,  Hily-Blant2019}, plotted as filled circles in the 2nd row 
and gray hatched bars in the 4th row, respectively.

\citet{Kastner2014} carried out a line survey around 300 GHz on the TW 
Hya disk using the 12 m Atacama pathfinder experiment (APEX) telescope. 
 The column densities of CN and C$_2$H derived from their hyperfine 
analyses were (9.6 $\pm$ 1.0) $\times 10^{13}$~cm$^{-2}$ and (5.1 $\pm$ 3.0) 
$\times 10^{15}$~cm$^{-2}$ respectively, when averaged within 5 
arcsec. 
 The HCN column density of (0.3 -- 9.3) $\times 10^{13}$~cm$^{-2}$ was 
derived under the assumption of local thermal equilibrium with an excitation 
temperature similar to those of the CN and C$_2$H lines (5~K -- 13~K). 

 ALMA observations and modeling with the HCN, CN, and C$_2$H lines was carried out 
for the TW Hya disk. 
 The CN and C$_2$H lines were observed to exhibit ring-like emission distributions with emission 
peaks around 40 and 60~AU, respectively \citep{Bergin2016,Cazzoletti2018,
Hily-Blant2017,Kastner2015,Nomura2016}. 
 In particular, \ector$>$~1 was required to reproduce bright C$_2$H emission 
rings \citep{Bergin2016}. 
The HC$^{15}$N ALMA observation indicated that the line intensity declines towards 
outer disk \citep{Hily-Blant2019}. 
 When gas temperatures of 30~K -- 50~K and gas density of 10$^{6}$ -- 
10$^{8}$~cm$^{-3}$ \citep{Bergin2016,Hily-Blant2019} were adopted for the  RADEX model
 \citep{vanderTak2007}, the observed fluxes of the optically thin lines  
 matched with the column densities of HCN, CN, and C$_2$H of
 10$^{14}$ -- 10$^{15}$~cm$^{-2}$ (see the black bars in the 1st row, the
gray hatched bars in the 3rd row, and the black bars in the 5th row, 
respectively, in Figure~\ref{fig:nc}). Here, a brightness temperature of 1 K was used for 
HC$^{15}$N 4-3 at 40~AU \citep{Hily-Blant2019}, the flux of 
0.166~Jy~km~s$^{-1}$ was integrated over two C$^{15}$N N=3-2, J=7/2$-$5/2 lines 
\citep{Hily-Blant2017} and the peak flux of 0.089 Jy~beam$^{-1}$ for the 
 C$_2$H N = 4 $-$ 3, J = 7/2 $-$ 5/2, F = 3 $-$ 3 line \citep{Bergin2016}, respectively.

 We note that the molecular column densities become higher under the assumption of
low gas density. 
 The HCN column densities at 40~AU are $\sim$10$^{14}$~cm$^{-2}$ and up to 
$\sim$10$^{15}$~cm$^{-2}$ when the gas density is higher than 10$^7$ cm$^{-3}$ 
and 10$^6$~cm$^{-3}$, respectively. The SMA observation of the HCN 3--2 line 
of the TW Hya disk showed that the HCN column density at 40~AU was 
$\sim$6$\times10^{13}$~cm$^{-2}$ \citep{Qi2008}. 
The HCN-abundant layer had a gas density higher than 10$^7$ cm$^{-3}$ 
and a gas temperature around 30 K in the model used in \citet{Qi2008}. 

The column densities constrain the initial gas-phase elemental abundances of 
carbon and oxygen. As shown in the bottom panels of Figure~\ref{fig:nc}, the 
C$_2$H column density is most sensitive to the \ector\, ratio and only those 
models having \ector$>$~1 can reproduce the observed column density 
reported by \citet{Bergin2016}.
 Furthermore, the models with \ector$>$~1 also reproduce the column density 
derived from the HC$^{15}$N ALMA observations \citep[see the black bars in the 
top panels of Figure~\ref{fig:nc}, ][]{Hily-Blant2019}. However, the observed CN 
column densities \citep[the dark gray horizontal bars in the middle rows of 
Figure~\ref{fig:nc}, ][]{Hily-Blant2017} are a factor of 10 lower than 
those from the models with \ector$>$~1.

 The ML and MS models were unable to reproduce the observed column density ratio of 
\hcnqcr, and thus the MH model with Equation~\ref{eq:frac} was introduced to satisfactorily
fit the observations.
In the observations \citep{Hily-Blant2018}, the column density ratio of \hcnqcr
 increases from $\sim$120 up to $\sim$330 as the distance from the central 
star increases from 20~AU to 50~AU. In our models with the ML and MS grain 
populations and \ector$>$~1, the column density ratios of \hcnqcr are roughly 
constant or decrease with the radius in the outer disk ($>$20~AU). 
The observed \hcnqcr ratio is similar to the ratio around 25~AU in 
the MLI3 model and to that around 40~AU in the MSI3 model. Therefore, we 
can infer that the hybrid model and the column density ratio of \hcnqcr in the MHI3 
model are consistent with the observations, as seen from 
Figure~\ref{fig:nc}.

 The MHI3 model could reproduce the observed ring-like emission distribution
of C$_2$H. The C$_2$H emission peaks at 60~AU and the central emission hole 
has an integrated intensity that is a factor of six weaker than that at the ring peak
 \citep{Bergin2016}. Their models with the spatially varied \ector\, ratio could 
 reproduce this ring-like emission \citep{Bergin2016}.
The C$_2$H column densities in the MLI3 and MSI3 models were spatially constant 
in the outer disk ($>$ 10~AU). On the other hand, the results of the MHI3 model showed that 
the column density of C$_2$H peaks around 40~AU and column density of C$_2$H 
at 40~AU is a factor of 10 higher than that in the inner disk ($<$ 20~AU). 
Therefore, the model with the spatially varied grain population could be 
another solution for the ring-like C$_2$H emission. However, it needs
additional fine-tuning.

\subsection{The best-fit model}
The grain population of the best fit model can be understood using grain 
evolution models. As mentioned above, the small dust grains in the atmosphere 
control the UV flux and \ncmn in the HCN-abundant 
layer, and change the \hcnqcr ratio. Therefore, the best fit model indicates 
that the small dust grains in the atmosphere of the outer disk ($>$ 20~AU) are 
depleted more than those of the inner disk. However, it is difficult to obtain 
the amount of small dust grains from the infrared observations of the (optically thick) 
light scattered off the dust grains toward TW Hya \citep[for example, ][]{Boekel2017}.
The small dust grains can be replenished when grain fragmentation is efficient and 
dominates the grain growth \citep[e.g.,][]{Dullemond2005,Birnstiel2012}. 
Generally, the inner disks are dominated by fragmentation, whereas the outer disks are 
growth-dominated and the small dust grains can survive for a few Myrs  
 in the case of strong turbulence \citep[e.g., ][]{Birnstiel2012}. 
 The grains become larger due to collisional sticking in the outer disk. 
They migrate to the inner disk and fragment into small dust grains if the 
turbulent motion is sufficiently strong.
Furthermore, this model can also explain the multiwavelength observations 
toward the TW Hya disk \citep{Menu2014}.

The \ector\, ratio also affects the \hcnqcr ratio. In this study, we assumed a 
spatially constant \ector\, ratio. 
 However, the infrared molecular line observations were interpreted using a model 
with a \ector\, of 0.5 within the disk radius of 2.4~AU \citep{Bosman2019}, 
whereas the molecular line observations from ALMA were explained using the models with 
\ector$>$~1 in the outer disk \citep[e.g., ][this work]{Bergin2016}.
Furthermore, models considering both grain and chemical evolution also 
showed that the \ector\, ratio varies in both radial and vertical directions 
\citep{Krijt2018,Krijt2020}. When the \ector\, ratio varies only along the 
distance from the central star, the column densities of HCN and C$_2$H 
change significantly \citep{Cleeves2018}. In the best fit model, the 
\hcnqcr ratio at 20~AU is slightly higher than the ratio obtained from experimental observations. When the 
\ector\, ratio increases with the height, the contribution of the NT layer to 
the total HCN column density can be increased. Then, the \hcnqcr ratio can 
be fitted satisfactorily. Furthermore, our model is unable to reproduce the column density 
and the nitrogen isotope ratio in CN, which might be solved by vertically 
varying the \ector\, ratio  and small \ector\, ratio in the surface layer of the disk.

\subsection{Effect of other parameters}
 We have investigated the effects of a few parameters on the results obtained from our model.
 The line-shielding effect has been included in this work and this effect could  
work in the disk atmosphere where small dust grains are depleted due to,
 for example, growth of the dust grains and their settling in the midplane of the disk. 
 The isotope-exchange reaction and the vibrationally excited 
H$_2$ also affect the \hcnqcr ratio. We ran the MHI3 model without the 
line-shielding effect, the isotope-exchange reaction, and the vibrationally 
excited H$_2$ in order to investigate their effect on the resulting molecular abundances 
and nitrogen fractionation. 
 For the X-ray ionization rate, we used the analytical model of \citet{Igea1999}.
However, depletion of small dust grains due to dust settling could reduce the X-ray opacity 
\citep{Bethell2011}, resulting in an increased X-ray ionization rate 
near the midplane of the disk. We also ran the model using the X-ray 
 ionization rate a factor of ten higher than that in the reference 
model. For the sake of simplicity, we assumed that the gas temperature is the same as that in the 
reference model, MHI3.  
The results thus obtained are plotted in Figure~\ref{fig:nc_check}.

The line-shielding effect increases the column densities of HCN, CN, and 
C$_2$H, but it does not change the column density ratio of the isotopologues 
significantly. When the line-shielding effect is considered, a larger number of molecules 
can survive the photodissociation process in the entire warm molecular layer. 
This effect is not significantly different for the NT layer and the 
LM layer, and thus the contributions of both layers to the total column density 
remain unchanged. Furthermore, the isotope ratios of HCN and CN in the NT layer 
exhibit a trend similar to atomic~N, which is 
affected by the \mn self-shielding effect and not by the line-shielding effect.
 
 The isotope-exchange reaction leads to only a minor change in the column density ratio of the
isotopologues. As mentioned above, we included the reaction 
$^{15}$N + CN $\leftrightarrow$ $^{14}$N + C$^{15}$N, which was ignored in the study by 
\citet{Visser2018}. However, the isotope-selective photodissociation of \mn 
predominantly affects the column density ratio of the isotopologues rather than 
the isotope-exchange reactions, as reported by \citet{Visser2018}, even if 
the abovementioned exchange reaction is included. 
However, in some cases, the isotope-exchange reaction contributes up to 60 \%: around 40~AU, the difference between the column density ratio of HCN from 
\enqcr in the reference model is 50, whereas that in the model without the 
abovementioned exchange reaction is 30 (it corresponds to the difference between 
the colored lines and the gray horizontal lines in the 2nd row in 
Figure~\ref{fig:nc_check}). 

The vibrationally excited H$_2$ increases the CN and HCN column densities 
and reduces the nitrogen isotope ratio of both molecules. The endothermic 
reaction between the vibrationally excited H$_2$ and atomic~N powers the 
formation of CN and HCN \citep[e.g.,][]{Visser2018}. This reaction is important
 in the NT layer, but not in the LM layer. Thus, when the vibrationally excited 
H$_2$ is considered, the contribution of the NT layer to the total column 
density increases and the column density ratio of the isotopologues is 
reduced further.

X-ray ionization is important in the LM layer and increases the column 
density ratios of the isotopologues as well as the column densities of CN and 
HCN. Higher ionization increases the abundances of assorted ions, resulting in 
increasing abundances of molecules other than CO and \mn \citep{Aikawa1999}. 
UV photons are the dominant ionization sources rather than X-rays in the NT layer 
(above $z/R>$~0.2), where the isotope-selective photodissociation of \mn 
reduces the nitrogen isotope ratio of atomic~N, CN, and HCN. Thus, in this 
layer, there is no difference between the models incorporating different X-ray 
 ionization rates. 
On the other hand, in the LM layer, X-ray photons form 
a larger number of ions, and thus larger quantities of CN and HCN. Thus, when the X-ray ionization rate increases, 
most CN and HCN molecules exist in the LM layer and both the column 
density and the isotope ratio increase.

\section{Summary}
\label{sec:summary}
We have developed a disk model for reproducing the nitrogen fractionation of the 
gas-phase HCN observed in the TW Hya disk. In the model, the isotope-exchange reactions and
isotope-selective photodissociation of \mn were included, which are important reactions 
that cause nitrogen fractionation. The effect of the grain population and 
initial abundances on the nitrogen fractionation process and the column density of HCN was investigated in this study.

The observed column density and the nitrogen isotope ratio of HCN could be 
reproduced when the \ector\, ratio was larger than unity and the small dust  
grains in the atmosphere were depleted considerably in the outer disk than in the inner 
disk. The \hcnqcr ratio was lower than 100 around the NT layer due to the 
isotope-selective photodissociation of \mn. The column density ratio of 
\hcnqcr depended on the relative contribution of the NT layer to the total HCN 
column density. The higher \ector\, ratio increased the HCN column 
density as well as the column density ratio of \hcnqcr because it led to 
a broader LM layer for HCN.  
The depletion of small dust grains in the atmosphere induced a narrower NT layer  
and a broader LM layer, which resulted in an increase the HCN column 
density and the column density ratio of \hcnqcr.
The radially constant grain population models showed that the column density
ratio of \hcnqcr is spatially constant or decreases outwards.
Thus, the observed \hcnqcr ratio can be 
explained by a model in which the small dust grains in the outer disk are
depleted more than those in the inner disk. This population of small 
dust grains is consistent with grain evolution models 
\citep[e.g., ][]{Birnstiel2012}. In this work, the line-shielding effects 
 for photodissociation by stellar UV photons were included, which enhanced 
the column density of HCN, but did not change the column density ratio of 
\hcnqcr.

\acknowledgments
{ We would like to thank the referee for comments that improved
our paper. This work is supported by the NAOJ leadership program,
MEXT Grants-in-Aid for Scientific Research 18H05441 and
19K03910, the Basic Science Research Program through 
the National Research Foundation of Korea
(grant no. NRF-2018R1A2B6003423), 
and NAOJ ALMA Scientific Research Grant Number of 2018-10B.
}

%%%%%%%%%%%%%%%%%%%%%%%%%%%%%%%%%%%%%%%%%%5
%
%     Reference
%
%%%%%%%%%%%%%%%%%%%%%%%%%%%%%%%%%%%%%%%%%%% 

\bibliography{Nitrogen_Fractionation_TWHya}
\bibliographystyle{aasjournal}

%%%%%%%%%%%%%%%%%%%
%
%  Tables & Figures
%
%%%%%%%%%%%%%%%%%%%%%

% Table1 {{{1 
\begin{table}
%\centering
\caption{Model parameters, and values for the reference model.}
\label{tab:table1}
\centering
\begin{tabular}{lcc}
\hline\hline

Quantity &Symbol &Value\\
\hline
stellar mass                      & $M_{\star}$      & $0.74\,M_\odot$\\
effective temperature             & $T_{\star}$      & $4110\,$K\\
stellar luminosity                & $L_{\star}$      & $0.28\,L_\odot$\\
UV luminosity                     & $L_{\rm UV}$     & $0.017\,L_\odot$\\
	X-ray luminosity                  & $L_X$           & $1.6 \times10^{30}\,{\rm erg\, s^{-1}}$\\
strength of interstellar UV       & $\chi^{\rm ISM}$ & 1 Draine field\\
cosmic ray H$_2$ ionization rate  & $\zeta_{\rm H_2}$
                                  & $\!\!1.0\times 10^{-19}$~s$^{-1}\!\!\!$\\
\hline
disk mass                   & $M_{\rm disk}$   & $0.01\,M_\odot$\\
gas/dust mass ratio         & $\delta_g$        & 100\\
concentration factor of the large dust grain & $\chi_d$    & 0.2 \\
inner disk radius                 & $R_{\rm in}$     & 0.07\,AU\\
outer disk radius                 & $R_{\rm out}$  & 200\,AU\\
reference and tapering-off radius  & $R_{c}$      & 150\,AU\\
column density power index        & $\gamma$      & 1\\
reference scale height            & $H_{\rm c}$     & 15\,AU\\
flaring power index               & $\beta$         & 1.3\\
\hline
\end{tabular}
\end{table}
%}}}1

%Table 2 {{{1 
\begin{table}[t!]
\caption{Initial abundances relative to the total hydrogen nuclei.}
\label{tab:initabun}
\centering
\begin{tabular}{cccc}
\hline\hline
Species & Abundance\tablenotemark{a} & Species & Abundance\tablenotemark{a} \\
\hline
H$_2$ & $5.000(-01)$   & He & $1.400(-01)$  \\
HCN\tablenotemark{b} & $1.000(-08)$     &  NH$_3$ ice\tablenotemark{b} & $9.900(-06)$ \\
N\tablenotemark{b} & $5.100(-06)$       &  \mn\tablenotemark{b} & $1.000(-06)$  \\
C & $7.000(-07)$       &  CH$_4$ & $1.000(-07)$  \\
CN\tablenotemark{b} & $6.600(-08)$      & H$_3^+$ & $1.000(-08)$ \\
HCO$^+$ & $9.000(-09)$ & C$_2$H & $8.000(-09)$ \\
C$^+$ & $1.000(-09)$   & Mg$^+$ & $1.000(-11)$ \\
Si$^+$ & $1.000(-11)$  & S$^+$ & $1.000(-11)$ \\
Fe$^+$ & $1.000(-11)$  &  \\
\hline\hline
Model   & \multicolumn{3}{c} {Species / Abundances\tablenotemark{a}} \\
\cline{2-4}
name    &   H$_2$O ice  & CO            & C \\
\hline          
I1      & $2.500(-04)$  &  $1.000(-04)$ & $7.000(-07)$\\
I2      & $2.500(-05)$  &  $1.000(-06)$ & $7.000(-07)$ \\
I3      & $2.500(-07)$  &  $1.000(-06)$ & $7.000(-07)$ \\
I4      & $2.500(-06)$  &  $1.000(-06)$ & $7.000(-06)$ \\
\hline
\end{tabular}
\tablenotetext{a}{ z(y) means z$\times$10$^y$.}
\tablenotetext{b}{ The initial abundance of $^{15}$N bearing species 
  is by a factor of \\ 330 lower than that of $^{14}$N bearing species except for \mn (165). }

\end{table}
%}}}1

%Table3 {{{1
\begin{table}
\centering
\caption{Free parameters in our models \label{tab:modellist}}
\begin{tabular}{cccccc}
\hline\hline
Name    & $f$\tablenotemark{a}  &  \multicolumn{4}{c}{Initial elemental abundances\tablenotemark{b}}   \\
\cline{3-6}
        &                        & IM\tablenotemark{c}   &    C       &  O   &    C/O   \\
\hline
MLI1    &     0.9                & I1   &  1.0(-4)   & 3.5(-4)  &  0.3   \\
MLI2    &     0.9                & I2   &  1.9(-6)   & 3.5(-6)  &  0.5   \\
MLI3    &     0.9                & I3   &  1.9(-6)   & 1.3(-6)  &  1.5   \\
MLI4    &     0.9                & I4   &  1.8(-5)   & 1.3(-5)  &  1.5   \\
\hline

MSI1    &     0.99               & I1   &  1.0(-4)   & 3.5(-4)  &  0.3   \\
MSI2    &     0.99               & I2   &  1.9(-6)   & 3.5(-6)  &  0.5   \\
MSI3    &     0.99               & I3   &  1.9(-6)   & 1.3(-6)  &  1.5   \\
MSI4    &     0.99               & I4   &  1.8(-5)   & 1.3(-5)  &  1.5   \\
\hline

MHI1    & Equation~\ref{eq:frac}  & I1   &  1.0(-4)   & 3.5(-4)  &  0.3   \\
MHI2    & Equation~\ref{eq:frac}  & I2   &  1.9(-6)   & 3.5(-6)  &  0.5   \\
MHI3    & Equation~\ref{eq:frac}  & I3   &  1.9(-6)   & 1.3(-6)  &  1.5   \\
MHI4    & Equation~\ref{eq:frac}  & I4   &  1.8(-5)   & 1.3(-5)  &  1.5   \\
\hline\hline
\end{tabular}
\tablenotetext{a}{Mass fraction of large dust grains.}
\tablenotetext{b}{z(y) means z$\times$10$^y$.}
\tablenotetext{c}{Model names in Table~\ref{tab:initabun}.}

\end{table}
%}}}1

 %Figures1-3{{{1
%Fig. 1

%Fig. 2
\begin{figure*}
\includegraphics[width=0.33 \textwidth]{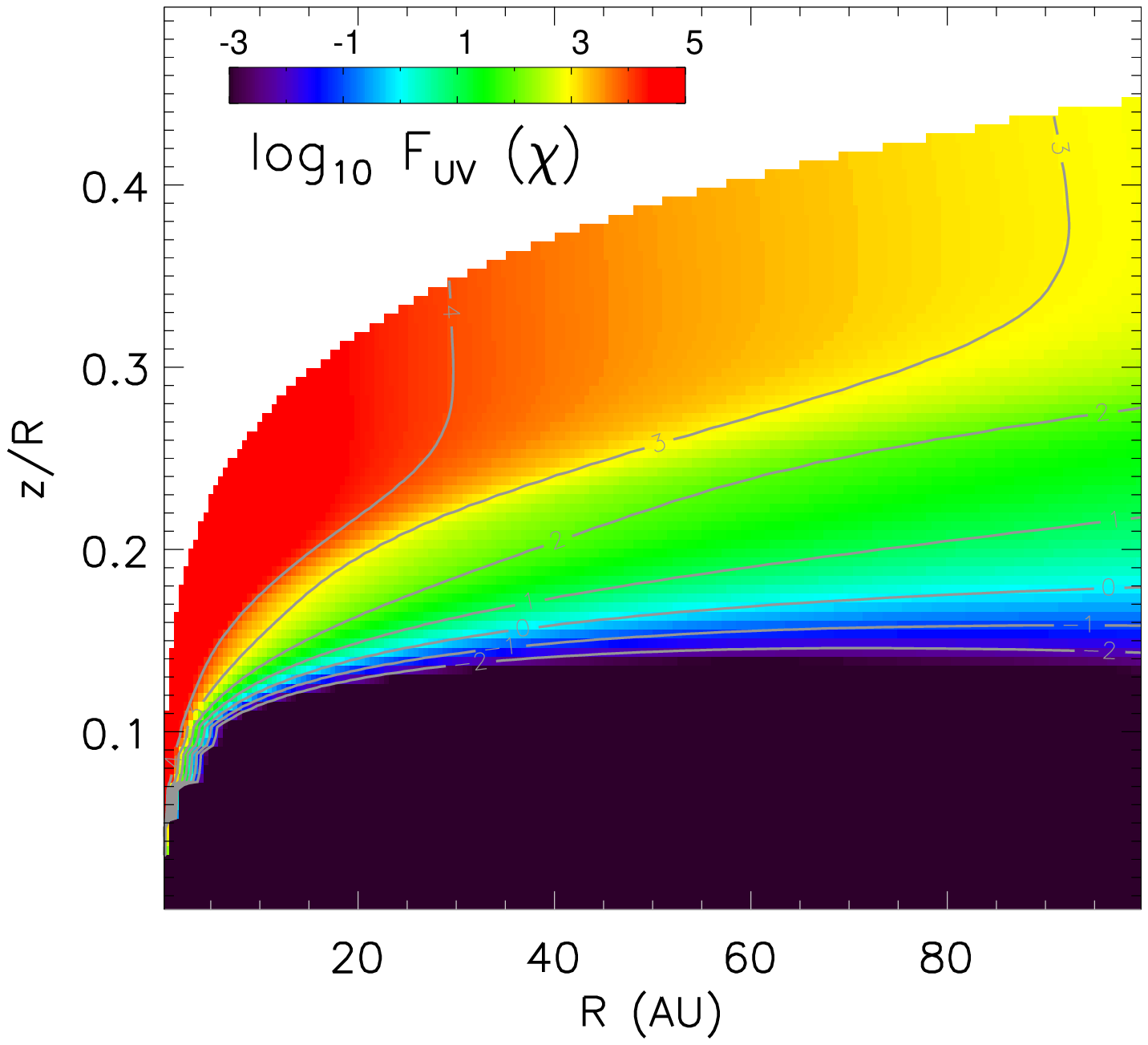}
\includegraphics[width=0.33 \textwidth]{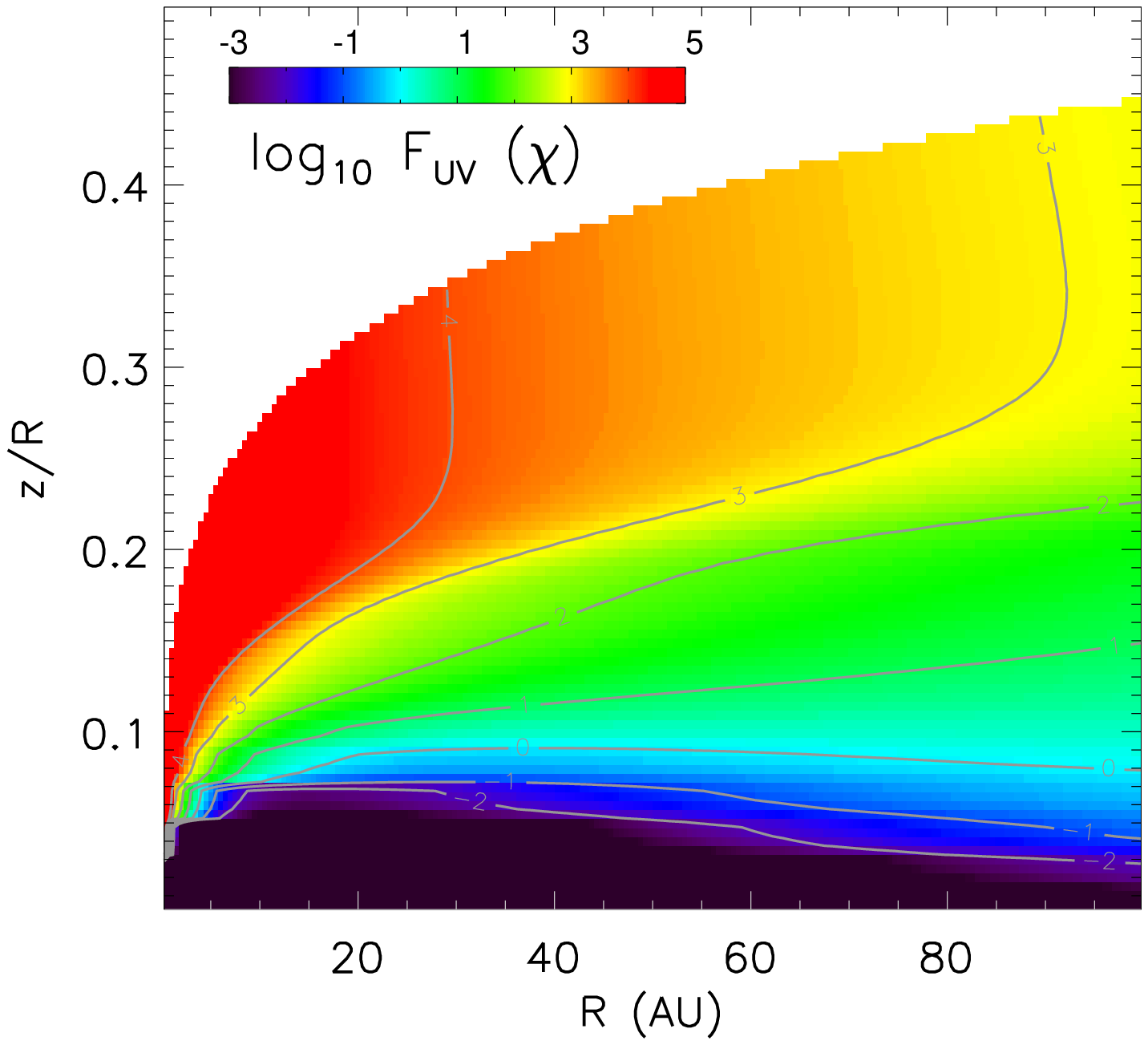}
\includegraphics[width=0.33 \textwidth]{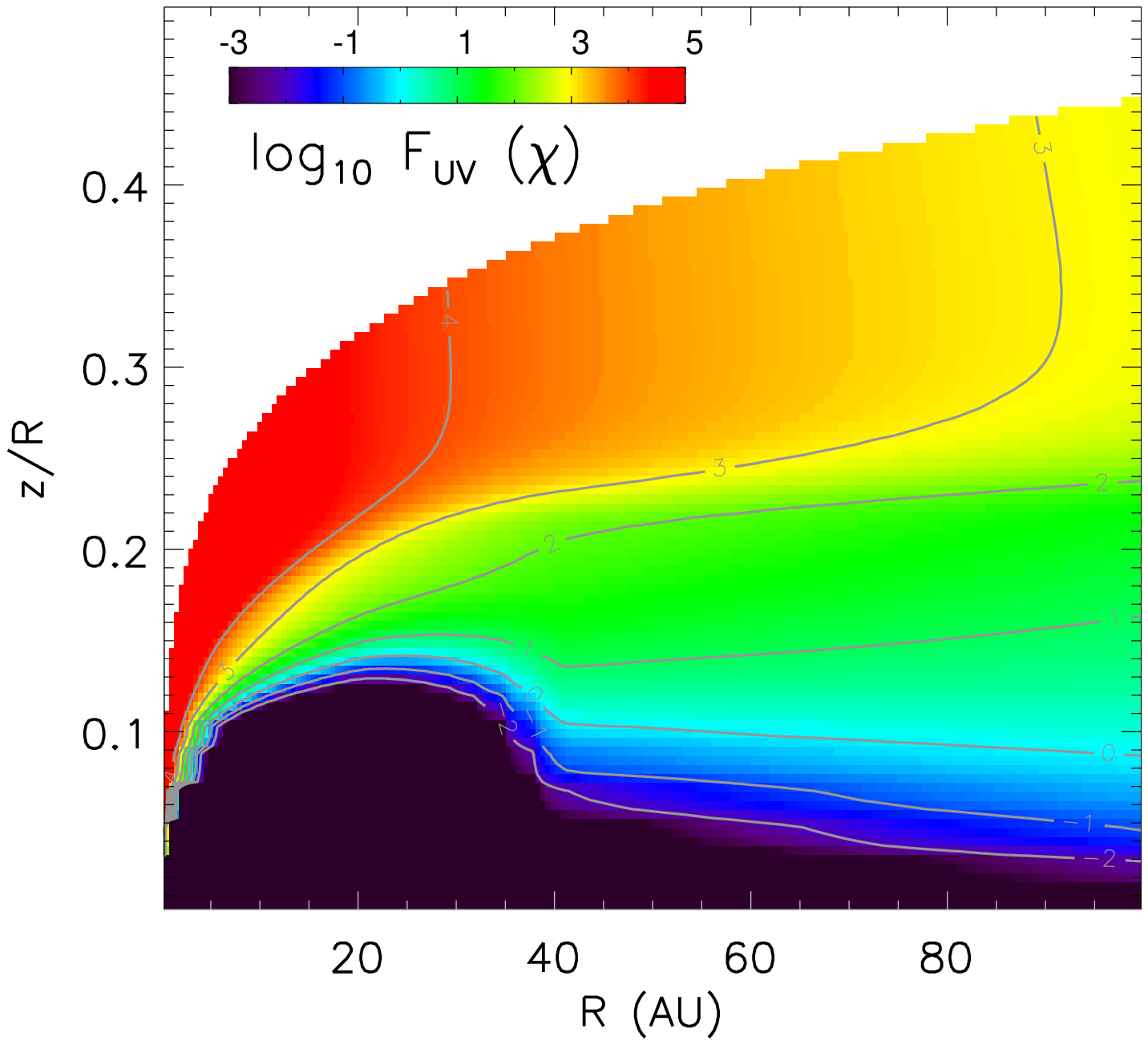}
\includegraphics[width=0.33 \textwidth]{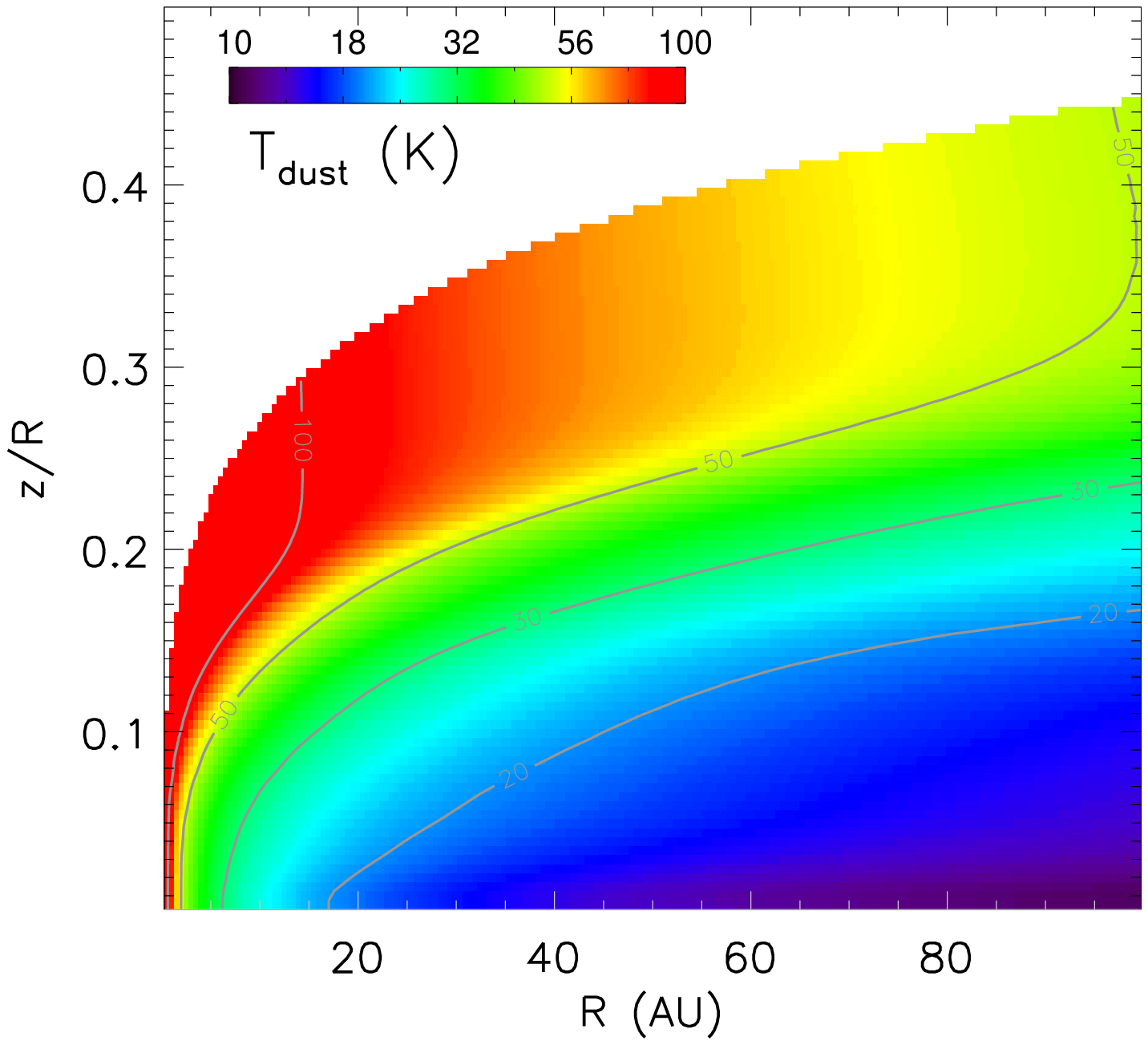}
\includegraphics[width=0.33 \textwidth]{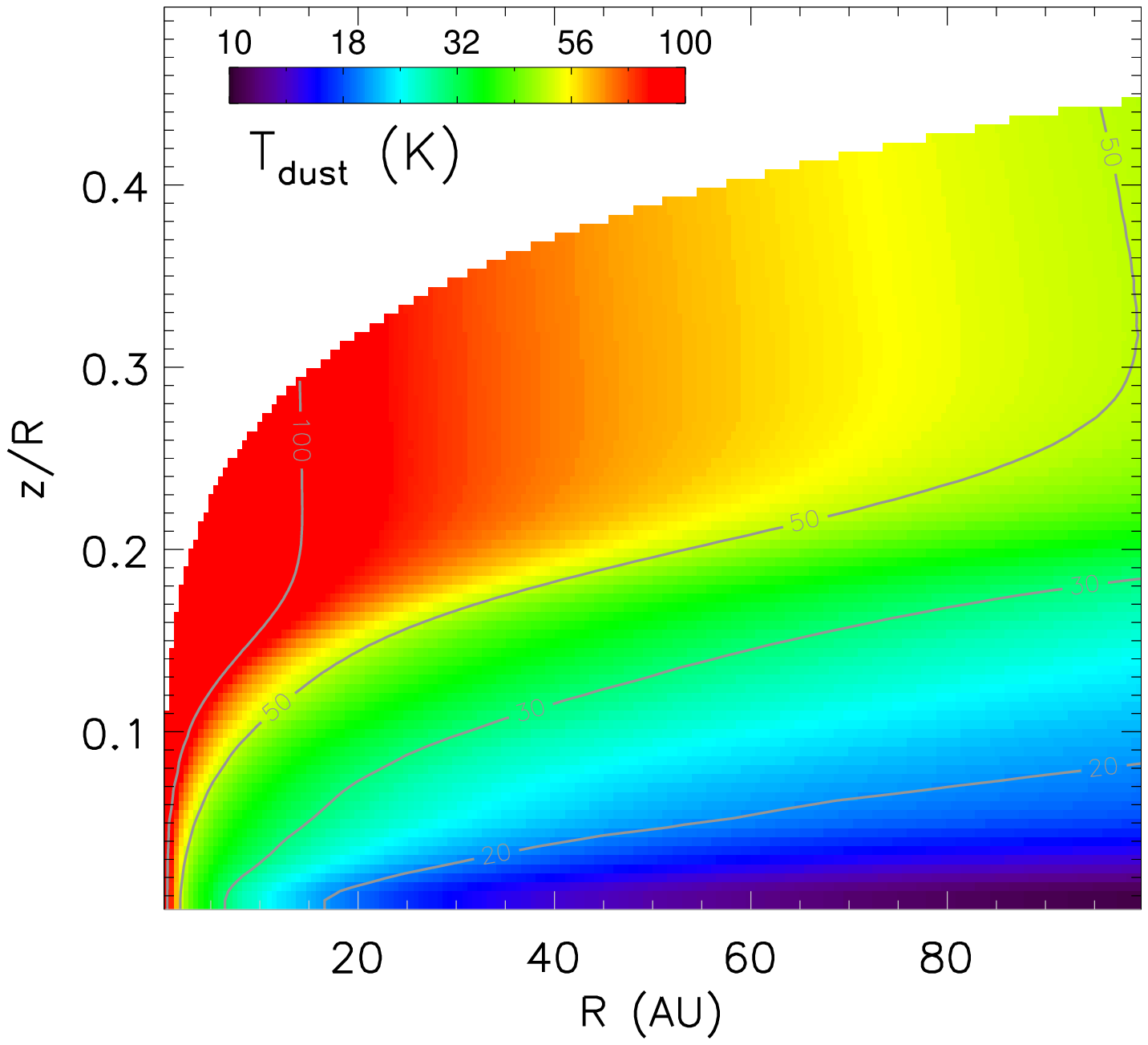}
\includegraphics[width=0.33 \textwidth]{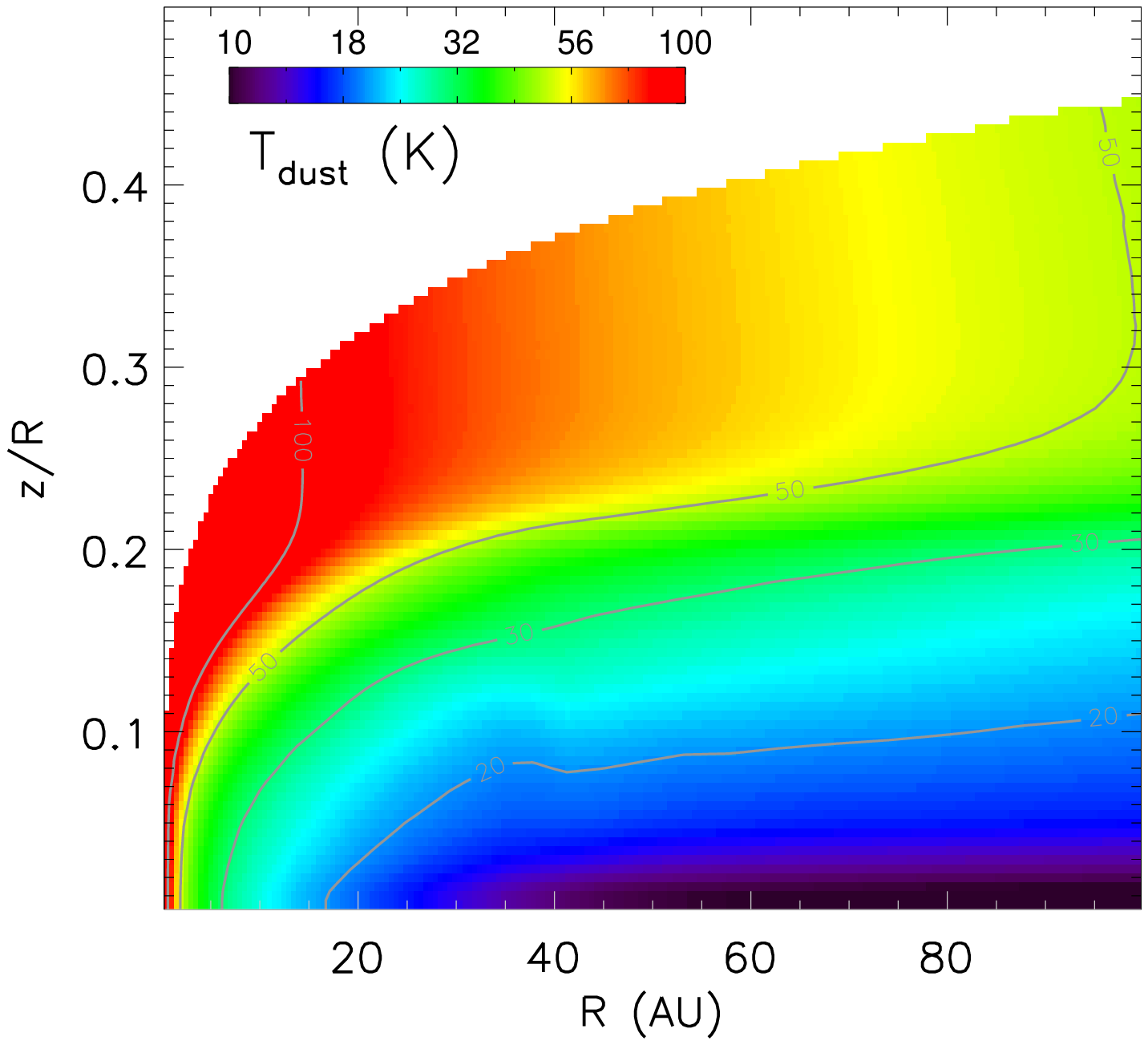}
\includegraphics[width=0.33 \textwidth]{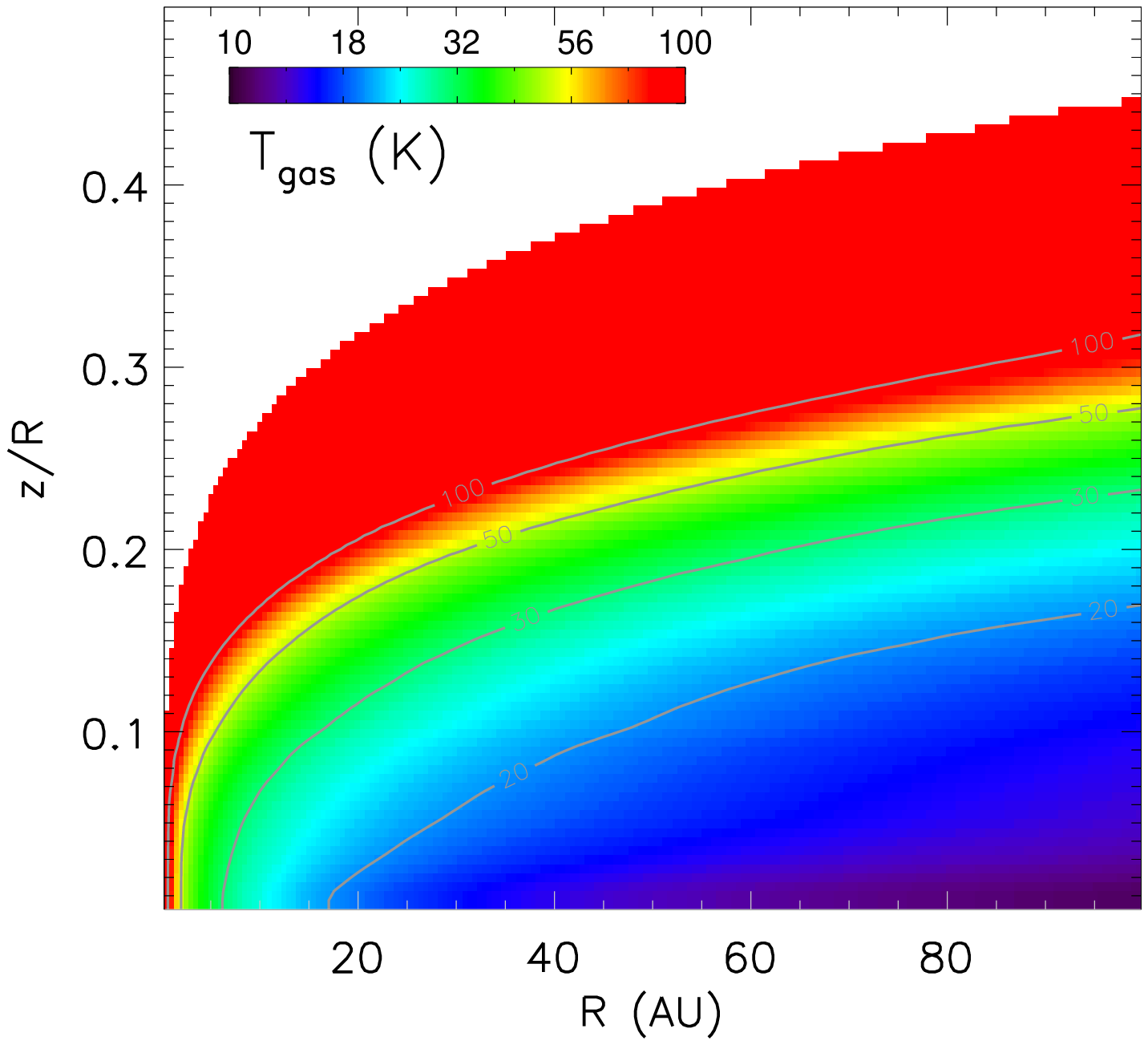}
\includegraphics[width=0.33 \textwidth]{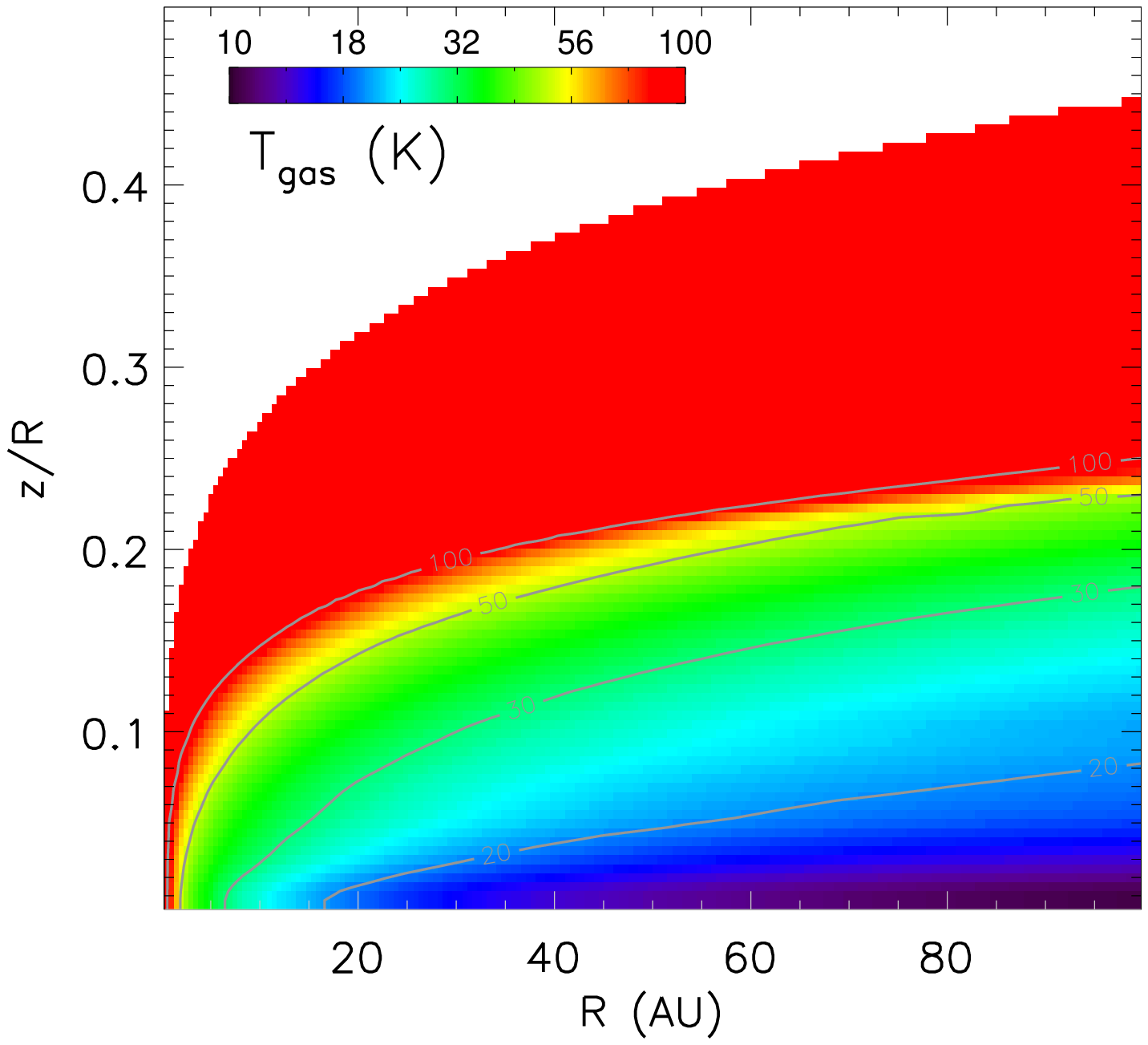}
\includegraphics[width=0.33 \textwidth]{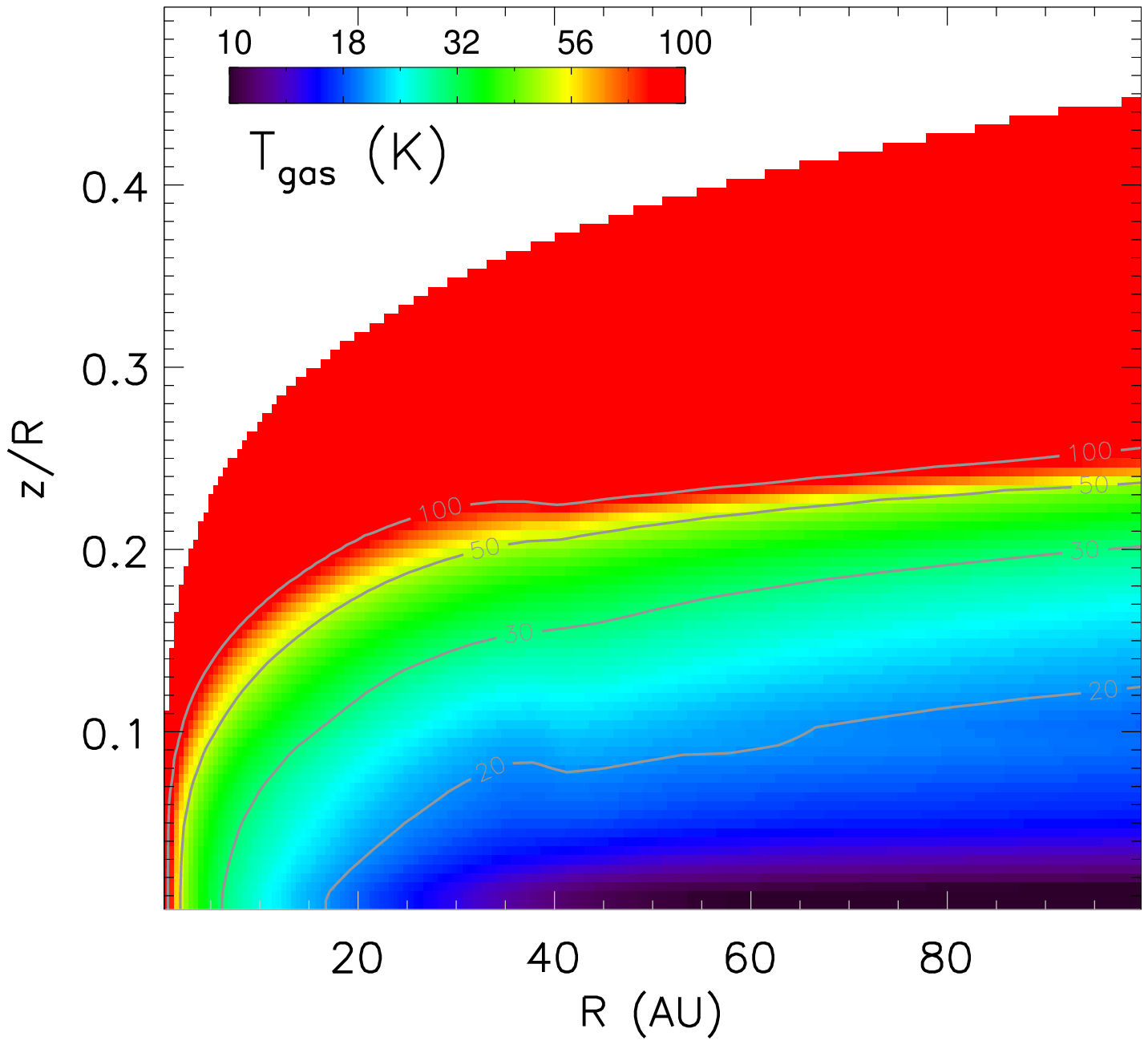}
\includegraphics[width=0.33 \textwidth]{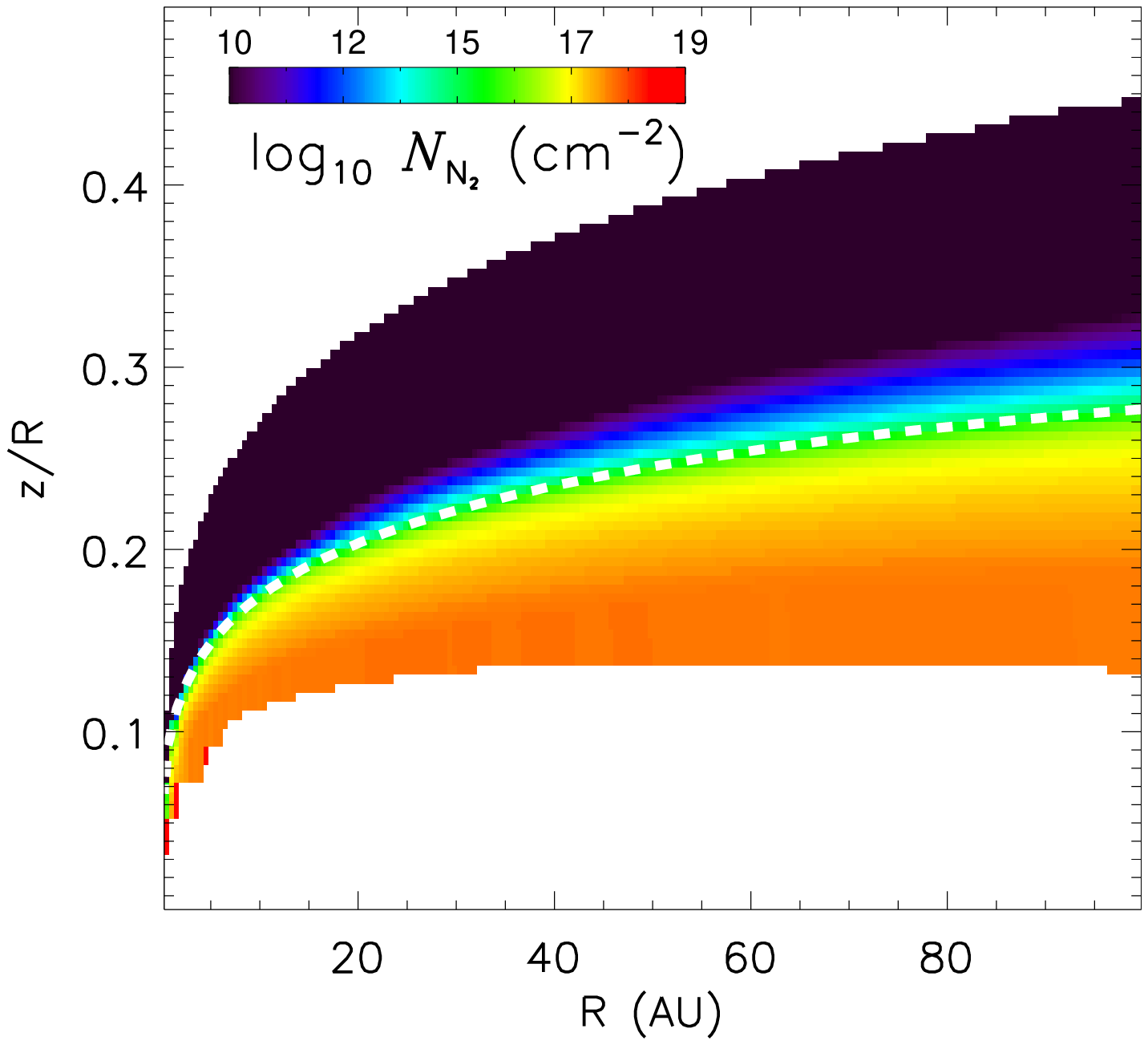}
\includegraphics[width=0.33 \textwidth]{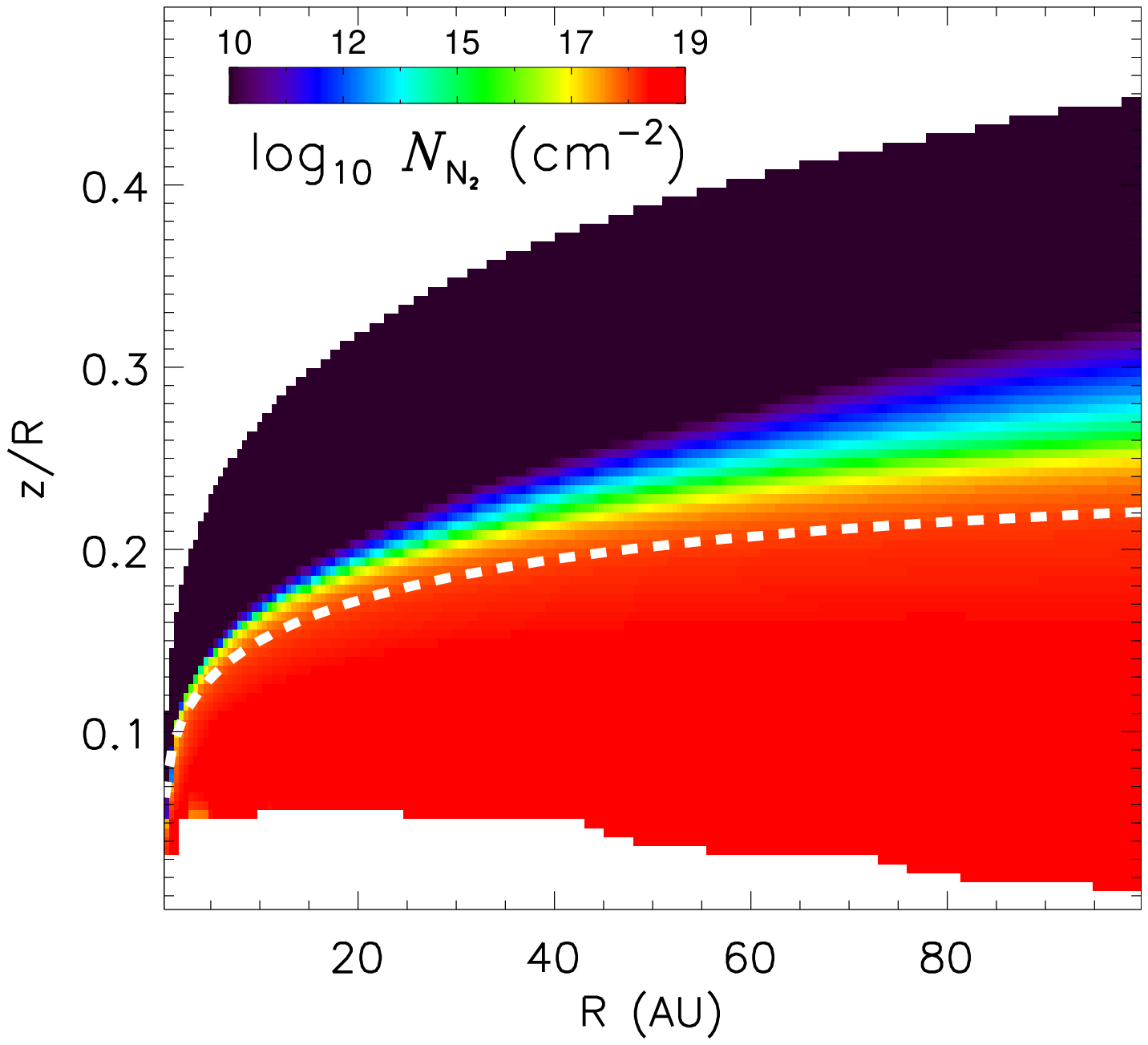}
\includegraphics[width=0.33 \textwidth]{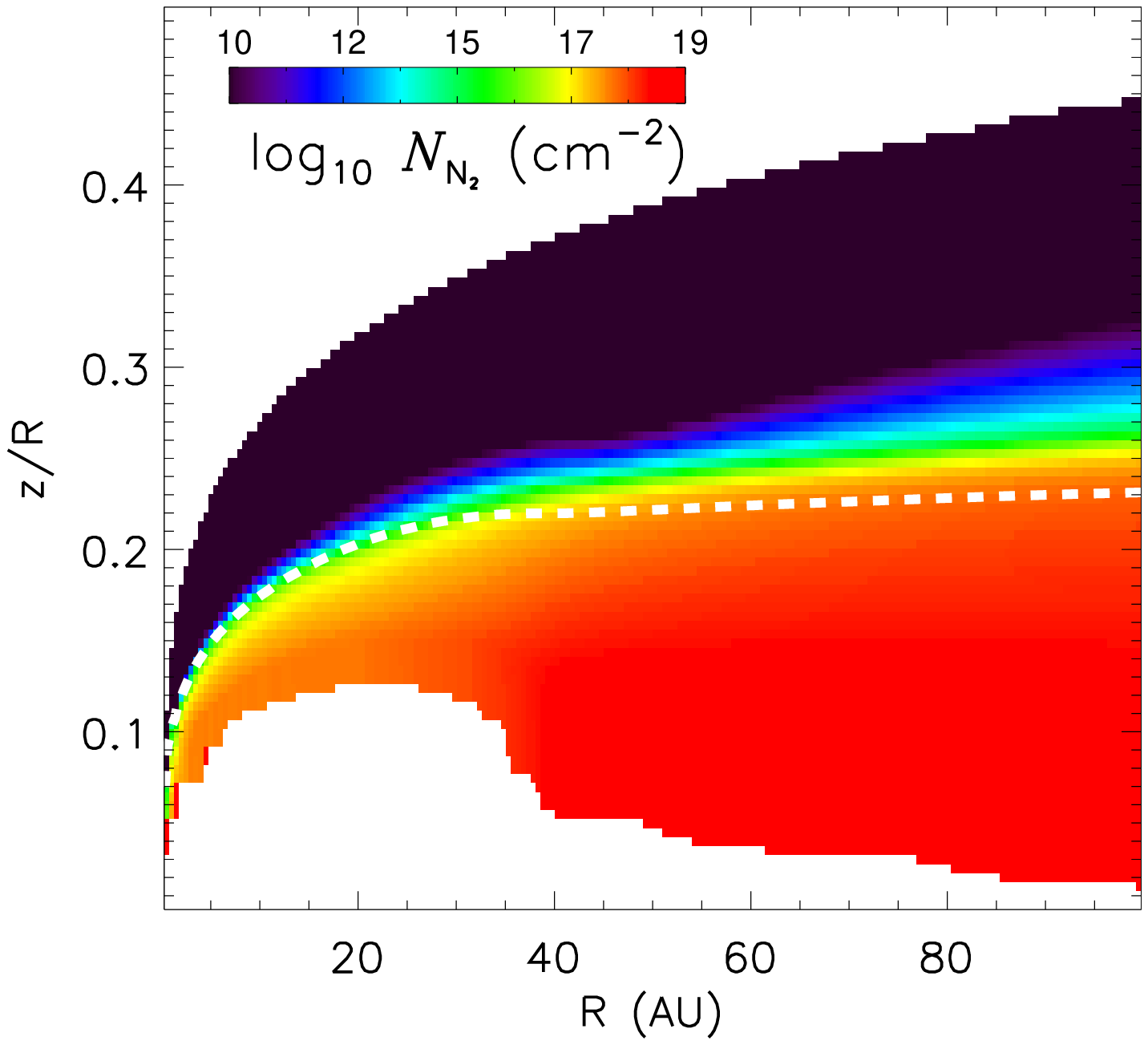}
\caption{UV flux (in the unit of Draine field), dust temperature, gas 
	temperature, and \mn column density (\ncmn) from the central star to a given position 
(from top to bottom) for the three grain 
population models: MLI3 ($f$=0.9; left), MSI3 ($f$=0.99; middle), and MHI3 
(see Equation~\ref{eq:frac}; right).
The white dotted lines in the bottom panels indicate the height where the 
 UV flux from the central star equals that from the upper atmosphere. 
Most UV photons directly originated from the central star above the 
white dotted line while the photons scattered by the dust grains are dominant in 
the disk atmosphere below the white dotted line.
}
\label{fig:model2}
\end{figure*}

%Fig. 3
\begin{figure*}
\includegraphics[width=1 \textwidth]{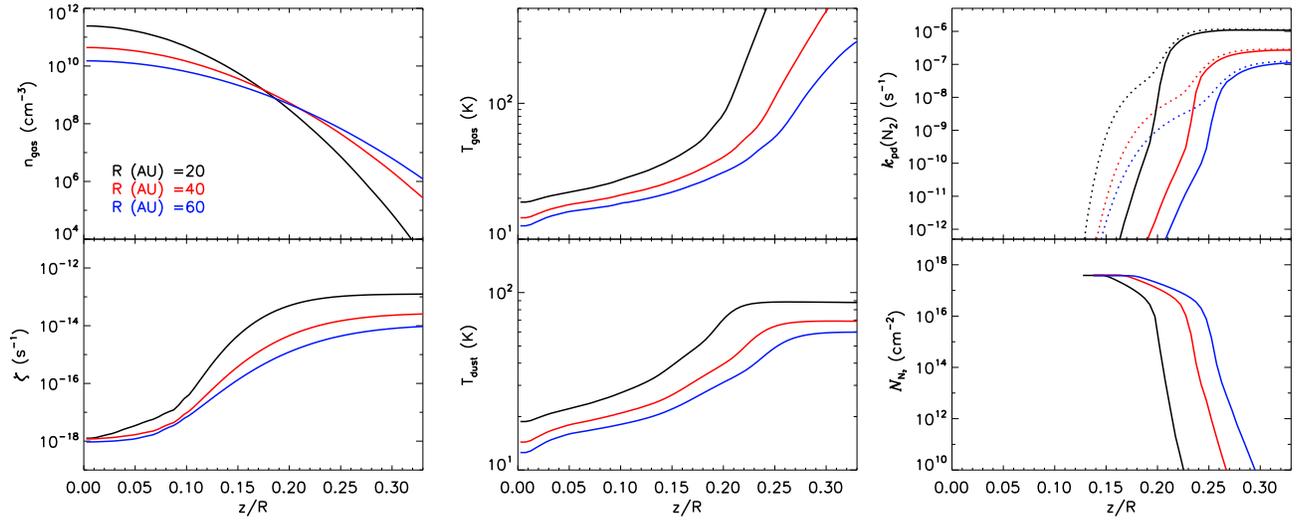}
\caption{ Vertical cuts of  Figures~\ref{fig:model} and \ref{fig:model2} 
for the MLI3 model at 20~AU (black), 40~AU (red), and  60~AU (blue). Top 
right panel: \mn photodissociation rates with (solid) and without (dotted) 
the self-shielding effect. 
}
\label{fig:model_v}
\end{figure*}
%}}}1

%Figures 4-6?{{{1
%Fig. 4
\begin{figure*}
\includegraphics[width=0.5 \textwidth]{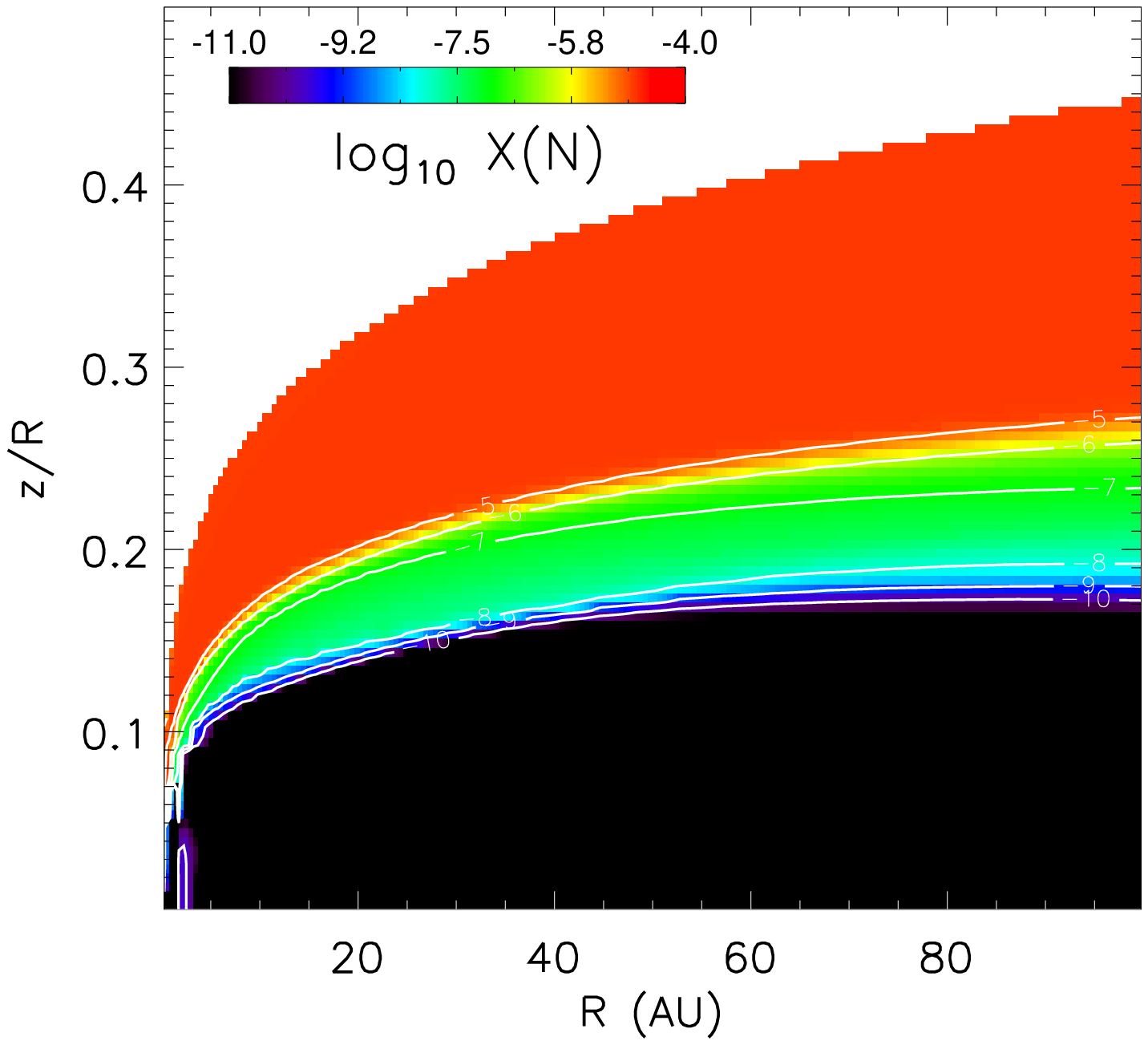}
\includegraphics[width=0.5 \textwidth]{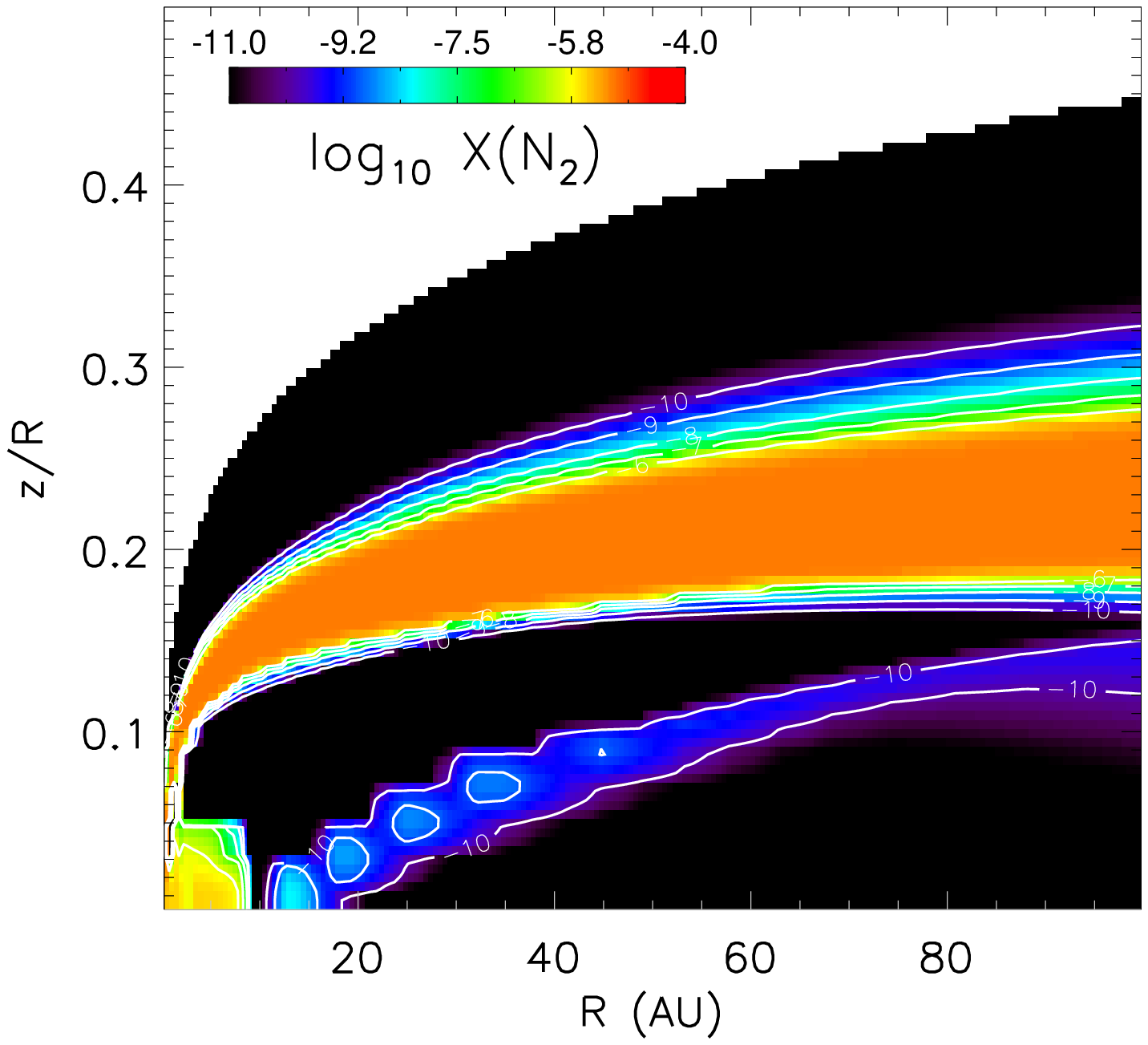}
\includegraphics[width=0.5 \textwidth]{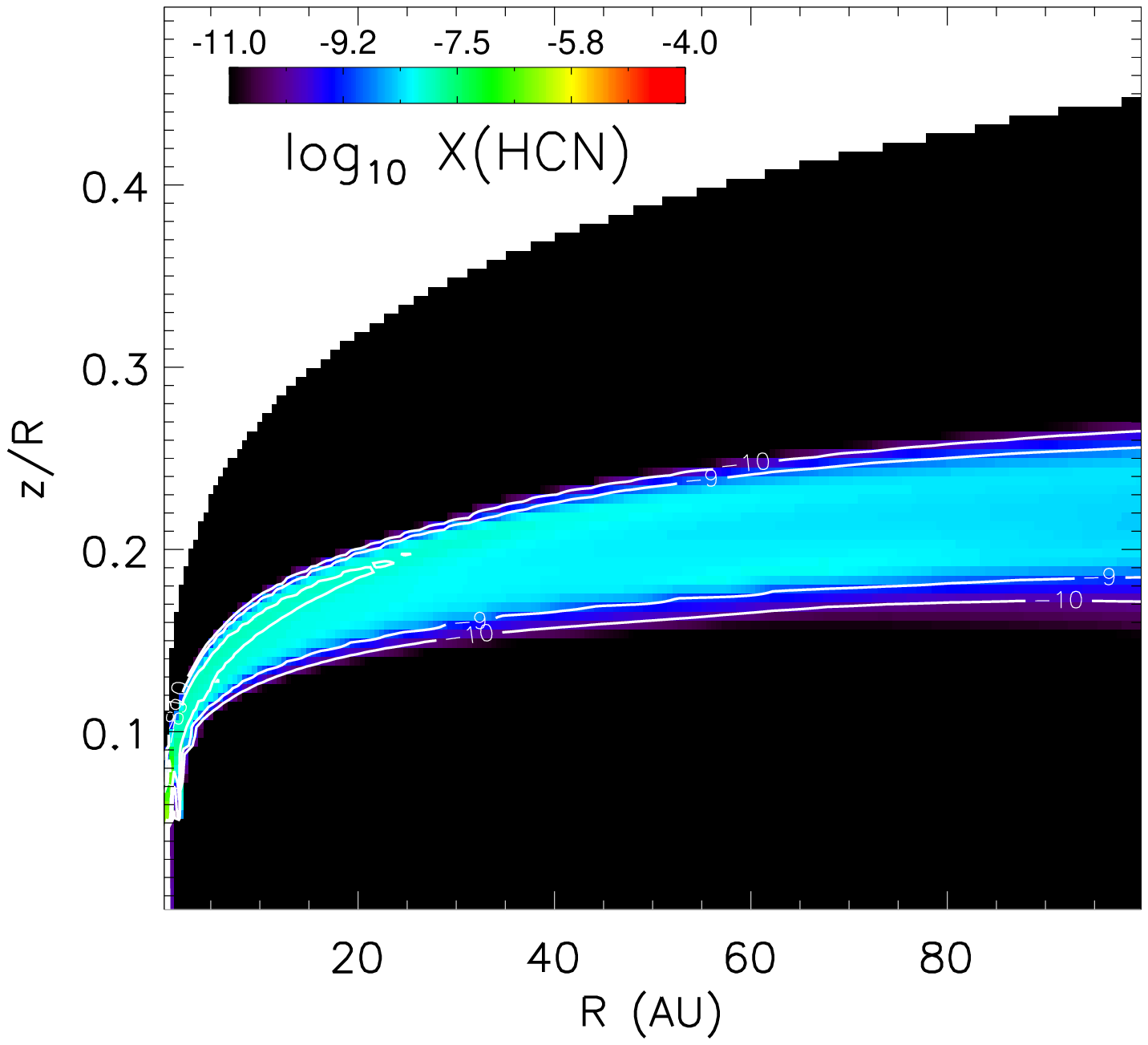}
\includegraphics[width=0.5 \textwidth]{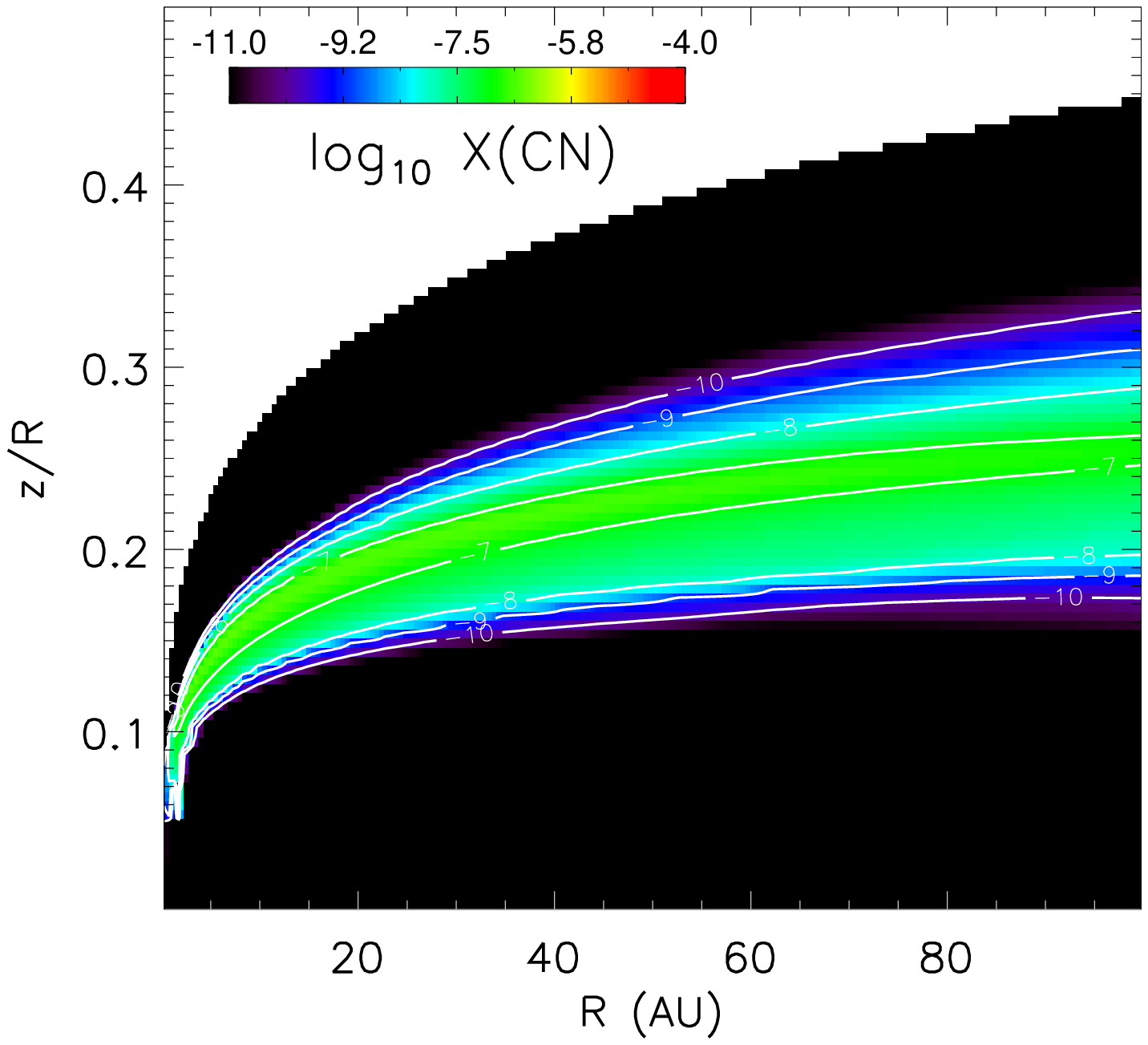}% crp
\caption{ Gas-phase abundance of N-bearing species for the MLI3 model. 
 The abundances relative to the total hydrogen nuclei in a log scale
are plotted in color image and contours. }
\label{fig:abun}
\end{figure*}

%Fig. 5
\begin{figure*}
\includegraphics[width=0.5 \textwidth]{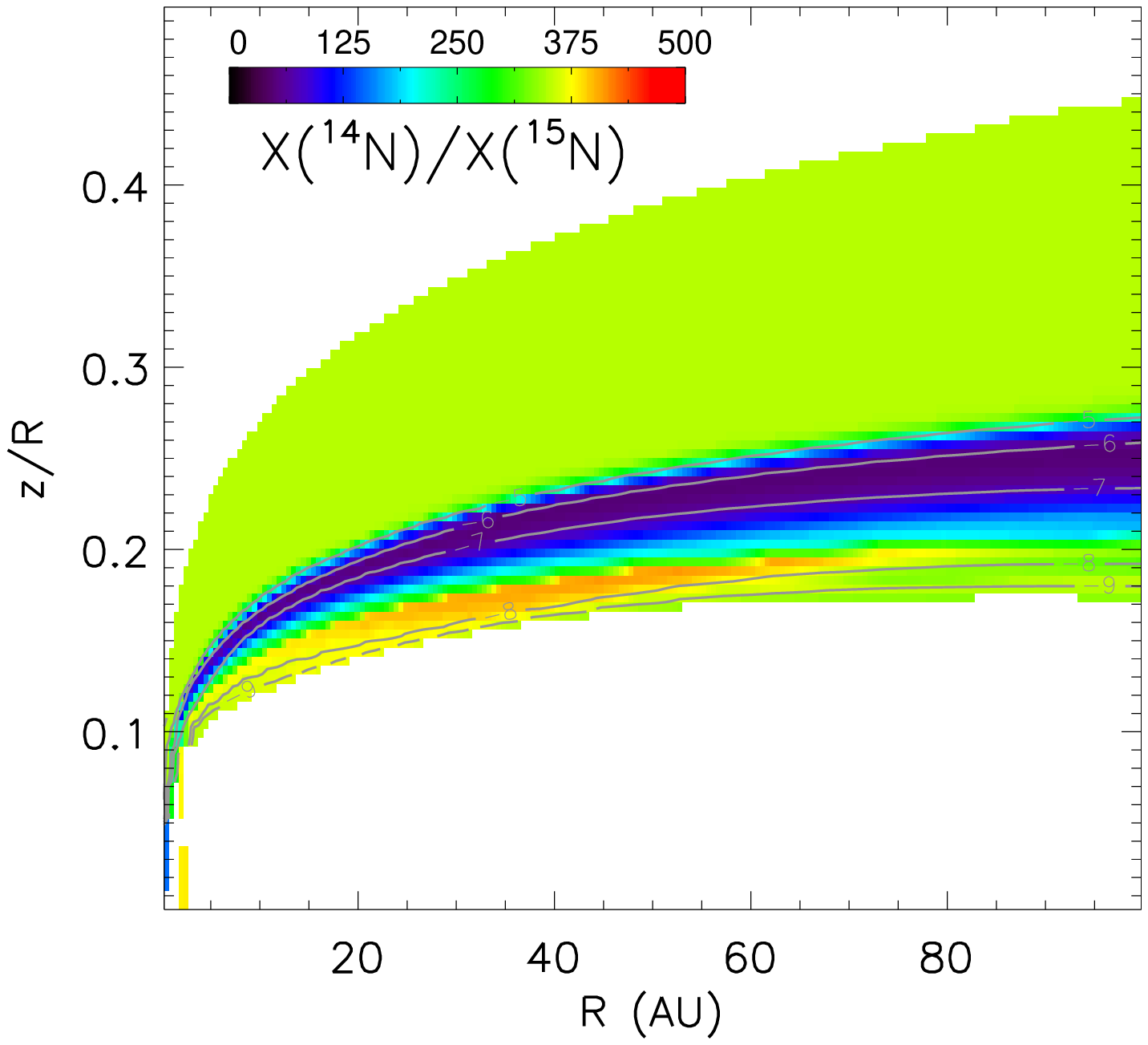}
\includegraphics[width=0.5 \textwidth]{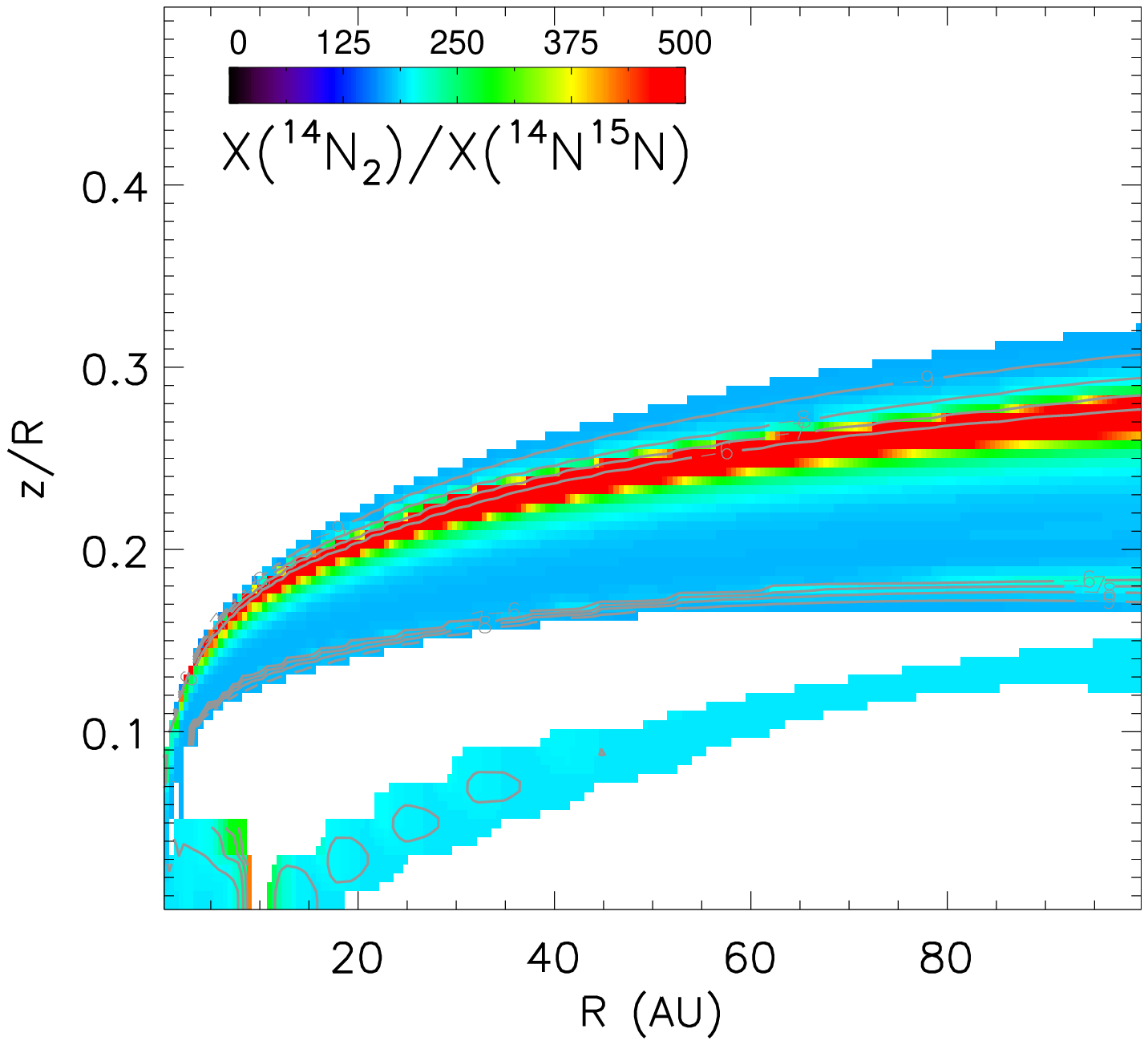}
\includegraphics[width=0.5 \textwidth]{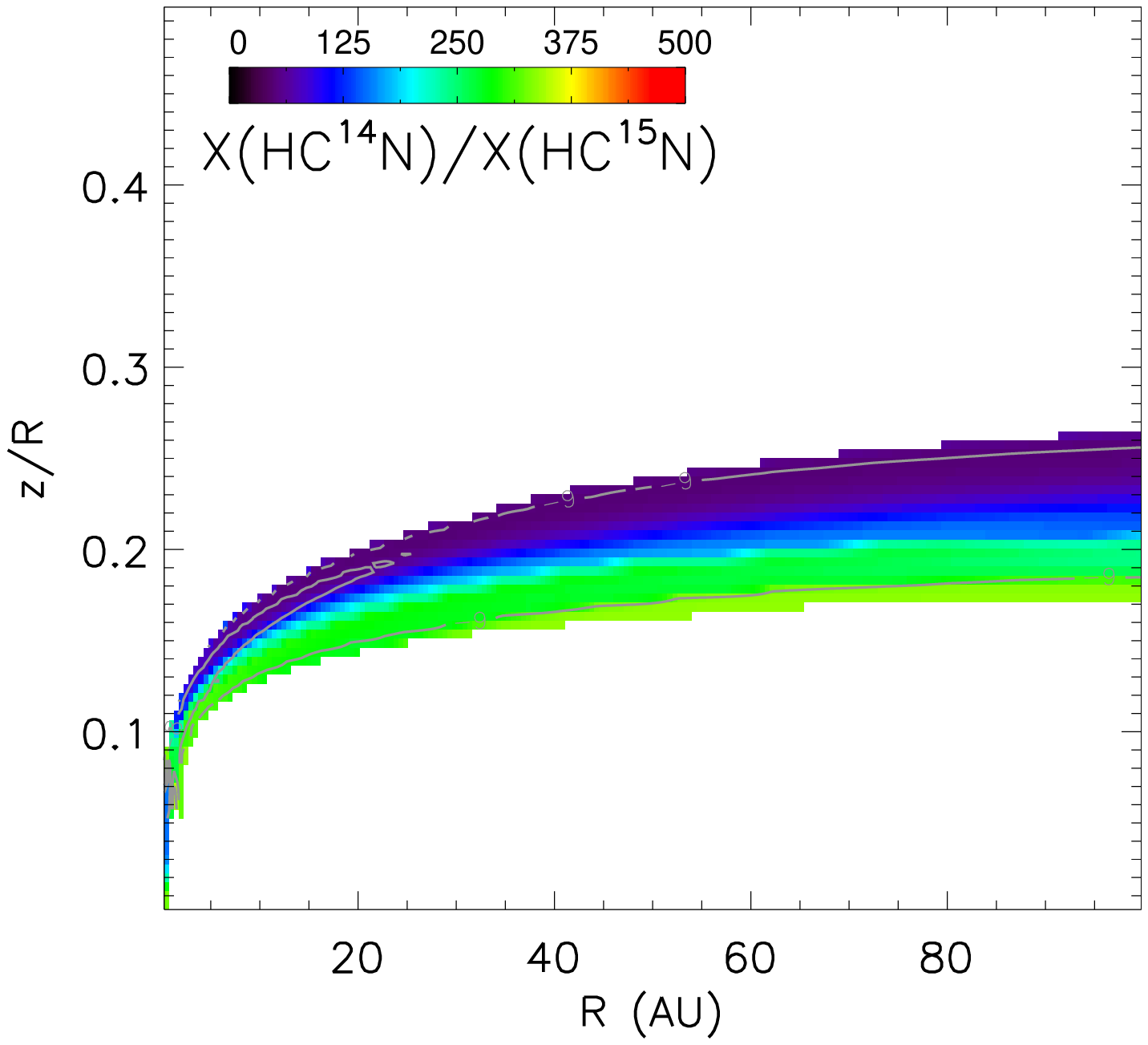}
\includegraphics[width=0.5 \textwidth]{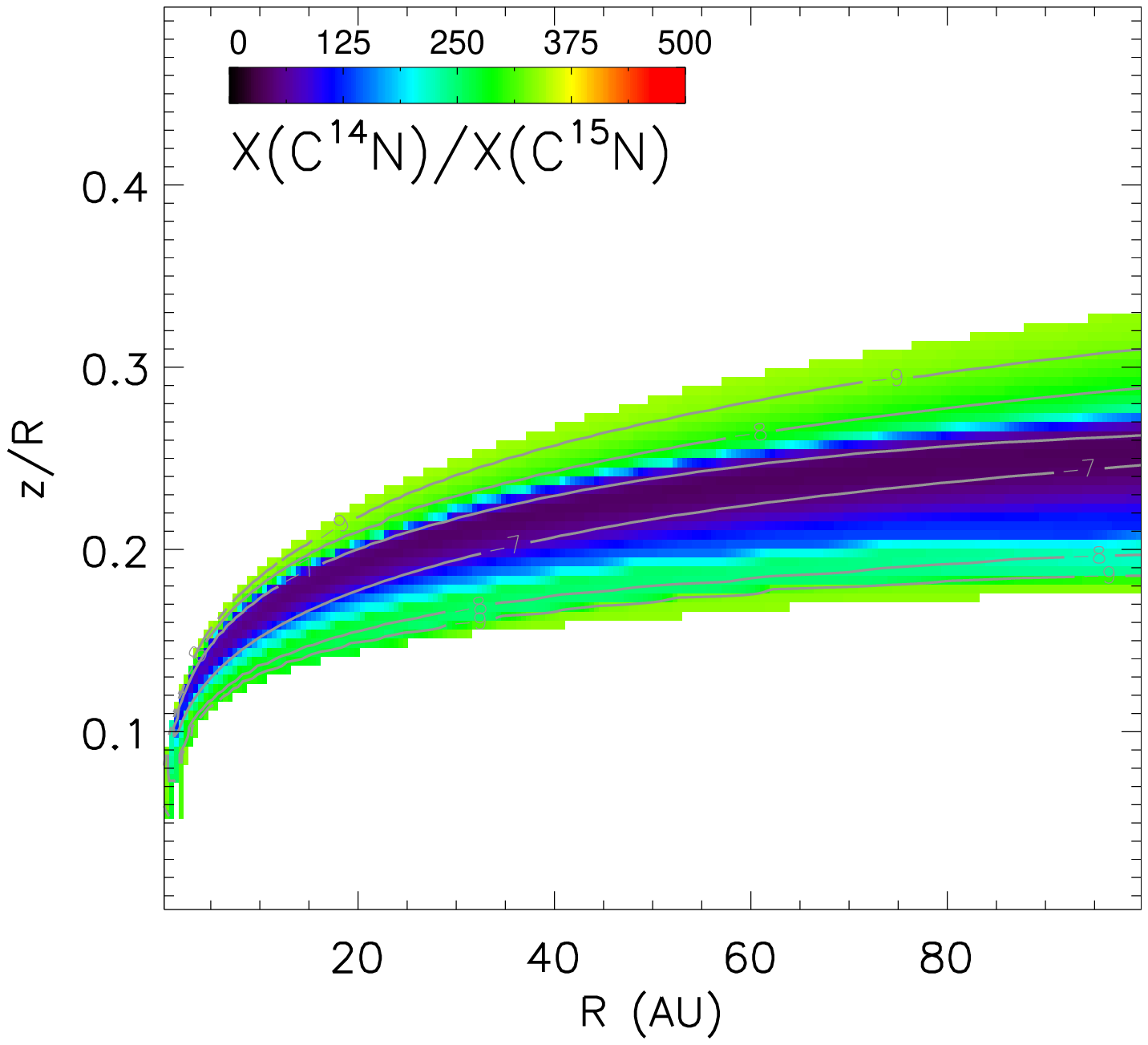}% crp
\caption{ Nitrogen isotope ratio for the species presented in 
Figure~\ref{fig:abun} for the MLI3 model. The contours indicate the 
abundances shown in Figure~\ref{fig:abun}. The regions where the abundance is 
lower than 10$^{-10}$ are masked and plotted in white. Note that the \enqcr 
ratio is 330 in our model.
}
\label{fig:isotope}
\end{figure*}

%Fig. 6
\begin{figure*}
\includegraphics[width=1 \textwidth]{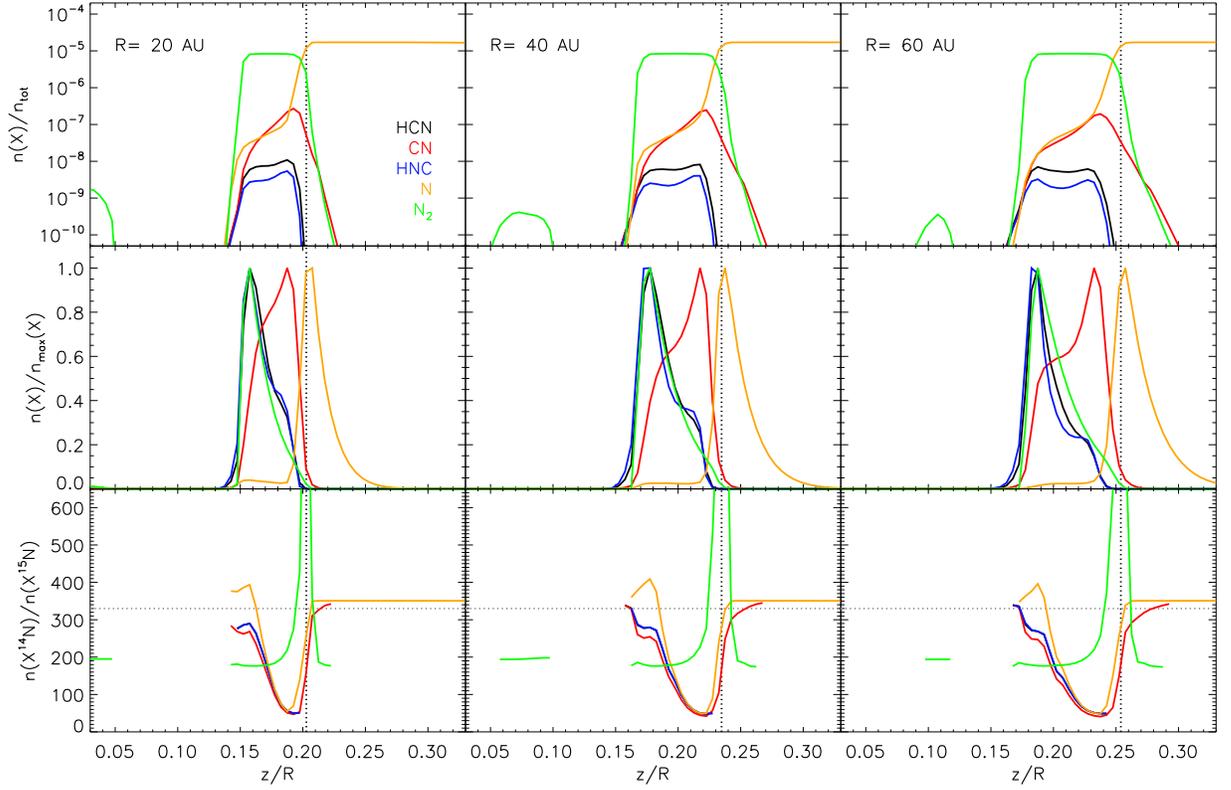}
\caption{Vertical cuts of  Figures~\ref{fig:abun} (top) and 
\ref{fig:isotope} (bottom) for the MLI3 model at 20~AU (left), 40~AU 
(middle), and 60~AU (right). The middle rows plot the 
number density normalized to the maximum number density of each species 
along the vertical direction in order to show the contribution to the 
column density. The dotted vertical lines indicate the height where the 
UV flux from the central star equals that from the upper atmosphere. \enqcr (=330) is presented in the dotted horizontal 
lines in the bottom panels.
}
\label{fig:abun_v}
\end{figure*}

\begin{figure*}
\includegraphics[width=1 \textwidth]{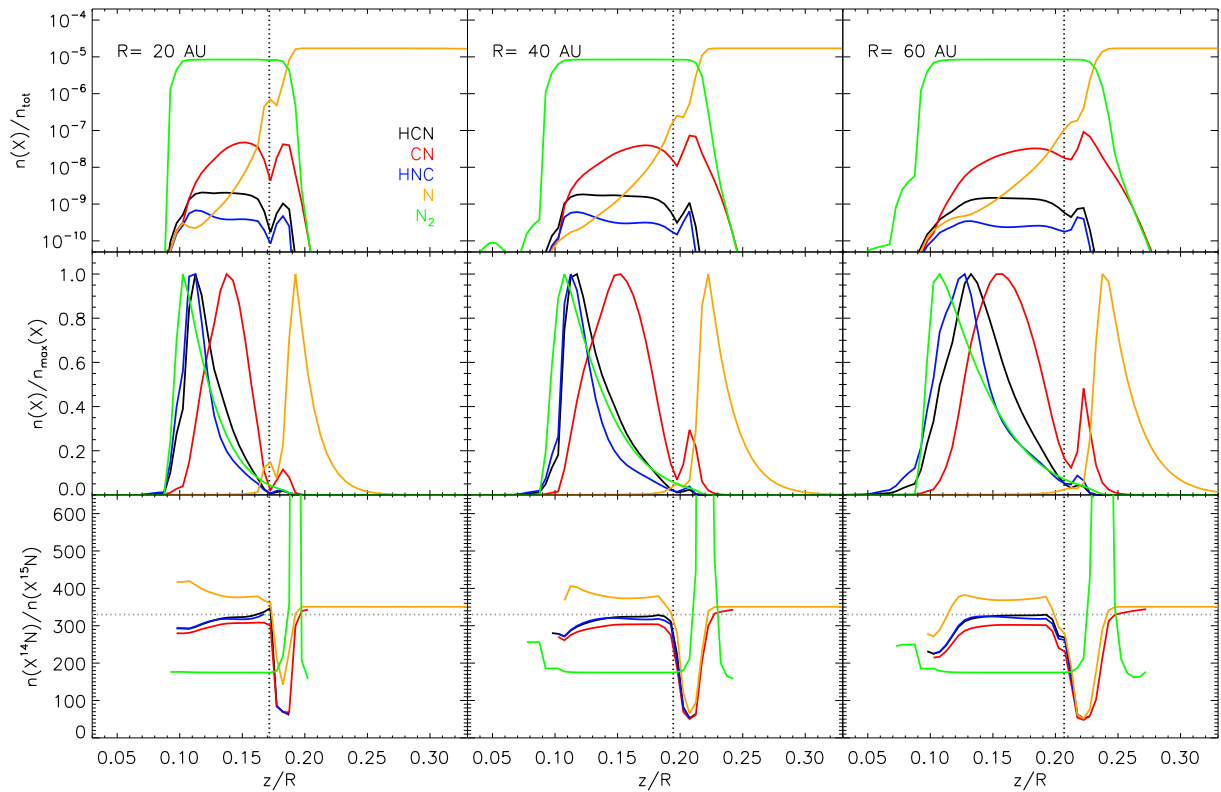}
\caption{Same as Figure~\ref{fig:abun_v} except for the MSI3 model.}
\label{fig:abun_v_s}
\end{figure*}
%}}}1

%Figures 8 {{{1
\begin{figure*}
\includegraphics[width=1 \textwidth]{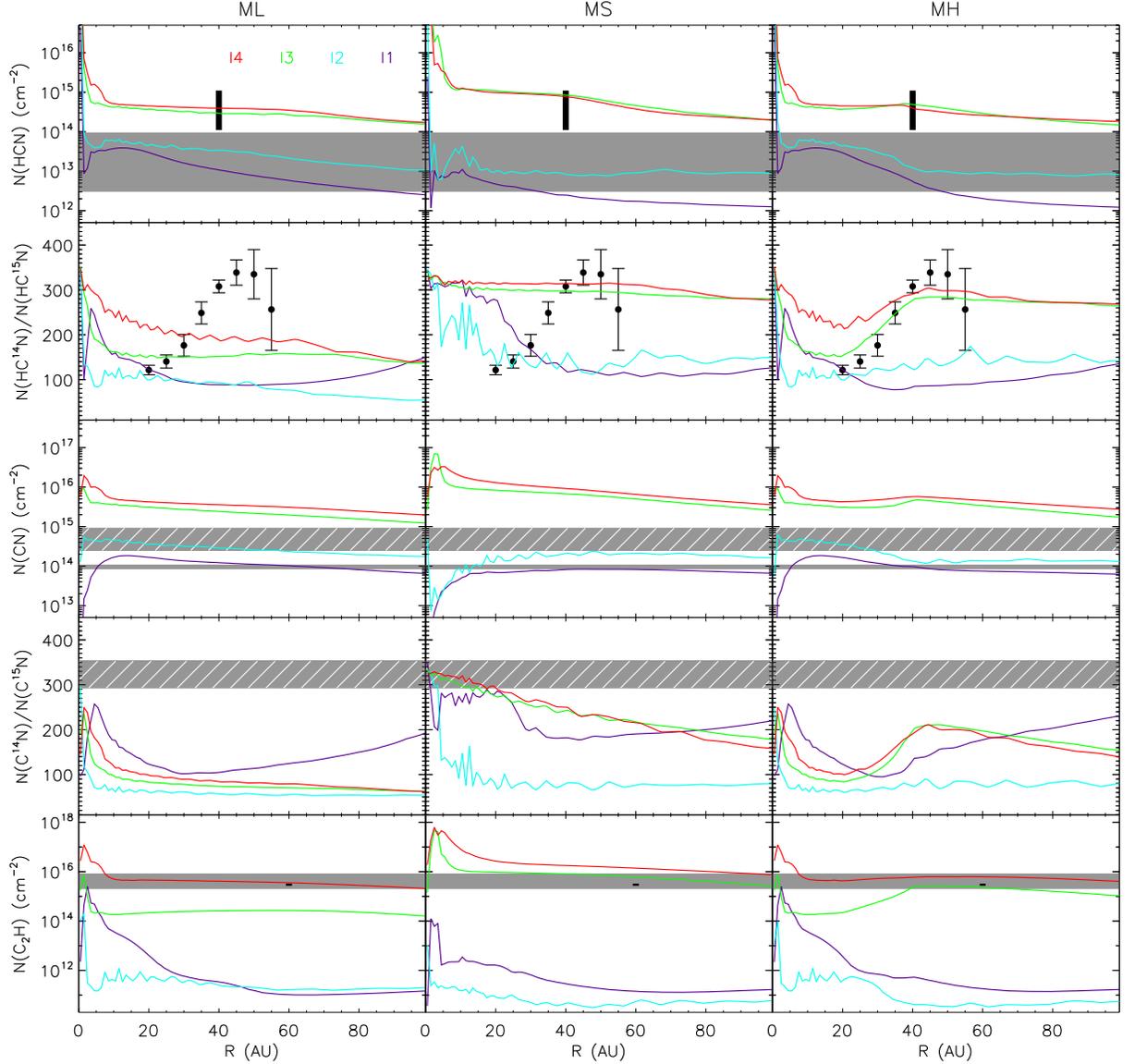}
\caption{HCN column density, column density ratio of 
HC$^{14}$N/HC$^{15}$N, CN column density, column density ratio of 
C$^{14}$N/C$^{15}$N, and C$_2$H column density (from top to bottom) for the 
ML (left), MS(center), and MH (right) models. The red, green, cyan, and purple 
lines indicate the result of models with different initial abundances of I4, 
I3, I2, and I1, respectively. The gray bars indicate the column 
densities derived from the APEX 12 m telescope observations \citep{Kastner2014}.
The column densities and column density ratios of isotopologue derived from the ALMA 
observations are presented in the black bars, filled circles, and gray hatched bars (see text).
}
\label{fig:nc}
\end{figure*}
%}}}1

%{{{1
\begin{figure*}
\includegraphics[width=0.3\textwidth]{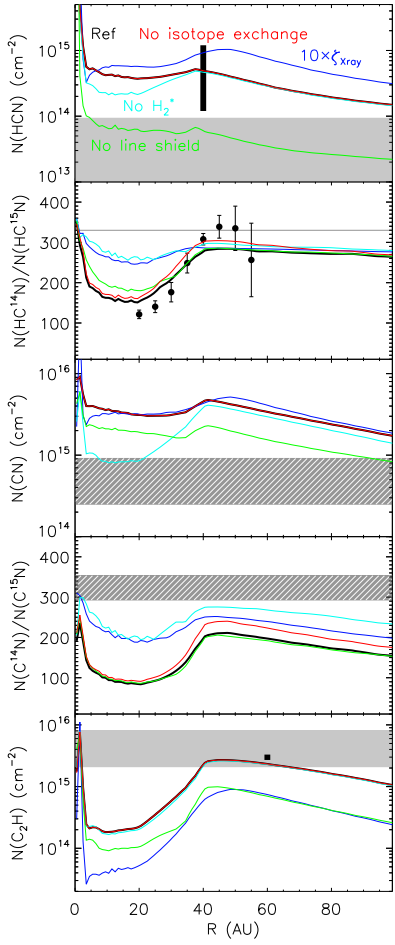}
\caption{HCN column density, column density ratio of 
HC$^{14}$N/HC$^{15}$N,  CN column density, column density ratio of 
C$^{14}$N/C$^{15}$N, and C$_2$H column density (from top to bottom) for the 
test models. Reference model is the MHI3 model (thick black line). The cyan, 
red, and green lines indicate the results for the models without the 
vibrationally excited H$_2$, the isotope-exchange reactions, and
the line-shielding effect, respectively. 
The blue lines represent the results for the model in which the X-ray 
ionization rate was a factor of 10 higher than that in the reference model. 
The observed data are the same as those in Figure~\ref{fig:nc}.
}
\label{fig:nc_check}
\end{figure*}
%}}}1
\clearpage
\newpage

\appendix
\section{Details of the PURE-C model}
\subsection{Dust grains}
\label{sec:dust}
The dust size distribution, $f_0(a)$ $\rm[cm^{-4}]$, was assumed to follow a power law 
of particle radius $a\,\rm[cm]$ as
\begin{equation}
  f_0(a)\;\propto\;a^{-3.5} 
  \quad\mbox{with}\quad
  a\in[a_{\rm min},a_{\rm max}] \ .
  \label{eq:dustsize}
\end{equation}
In this work, we used two grain populations: small dust grains with radii  
$r_g$= 0.005~$\mu$m -- 1~$\mu$m and large dust grains with radii $r_g$= 0.005~$\mu$m 
-- 1~mm.
The representative radii ($a_{\rm dust}$) of the small and large dust grains 
are 0.1 $\mu$m and 10~$\mu$m following the method given by \citet{Vasyunin2011}.

The dust opacity was calculated using the OPACITY-TOOL\footnote{
\url{https://dianaproject.wp.st-andrews.ac.uk/data-results-downloads/fortran-package/}}
with the discrete dipole approximation \citep{Min2016} and the dust grains were considered to be a mixture of 
amorphous laboratory silicates \citep[][$\rm Mg_{0.7}Fe_{0.3}SiO_3$]{Dorschner1995} 
with amorphous carbon \citep[][BE-sample]{Zubko1996}.  
We adopted the reference parameters for the dust grains as given in \citet{Woitke2016}: 
dust material density of 2.09 g~cm$^{-3}$, volume fractions of silicate, 
amorphous carbon, and porosity of 60 \%, 15 \%, and 25 \% respectively, and 
the maximum hollow volume ratio of 0.8.

\subsection{Dust continuum radiative transfer} \label{subsec:dcrt}
%{{{1
The dust continuum radiative transfer problem was solved using a Monte Carlo 
method described in the article by \citet{Lucy1999}. The dust temperature was calculated 
by assuming a radiative equilibrium:
\begin{equation}
\label{eq:re}
4 \pi \int_0^\infty \kappa_\nu B_\nu (T) d\nu 
= 4 \pi \int_0^\infty \kappa_\nu J_\nu d\nu,
\end{equation}
where $\kappa_{\nu}$ is the absorption coefficient of the dust grain at the 
frequency $\nu$, $B_\nu (T)$ is the Planck function with the dust temperature T, 
and $J_\nu$ is the mean intensity in the cell.
For a given dust grain, the left-hand side of Equation~\ref{eq:re} is the total 
emission from the dust grain in the cell with the dust temperature.
The right-hand side is the total energy absorbed by the dust grain and is derived 
 numerically by
  \begin{eqnarray}
 4 \pi \int_0^\infty {\kappa_\nu J_\nu d\nu} = \frac{L_{\rm tot}/N_\gamma}{V} \sum_i \Delta \tau_i 
 \end{eqnarray}
 where $L_{\rm tot}$ is the total luminosity in the system, $N_\gamma$ is the 
number of model photons, $V$ is the volume of the cell, and $\Delta \tau_i$ is 
the optical depth of the trajectory of a model photon $i$ within the grid cell. 
The dust temperature is converged within a few iterations. 
  
In the very optically thick midplane of the disk, where the Monte Carlo method 
was not efficient, a modified random walk method \citep{Min2009,Robitaille2010} 
was applied. 
When a photon escapes from the optically thick and homogeneous sphere of 
radius R and dust grain density $\rho_{\rm dust}$, the dust grain within 
the sphere absorbs an energy per unit dust grain density of
\begin{equation}
E = (L_{\rm tot}/N_\gamma) \kappa_{\rm P} \left(\frac{R^2}{2}\overline{\chi^{-1}}\rho_{\rm dust}\right)
\end{equation}
 where 
 \begin{equation}
 \label{eq:kp}
	 \kappa_{\rm P} =\int_0^\infty \kappa_{\nu} B_\nu (T) d\nu  / \int_0^\infty B_\nu (T) d\nu 
 \end{equation}
 \begin{equation}
 \label{eq:chi}
   \overline{\chi^{-1}}=\int_0^\infty \chi_\nu^{-1} B_\nu (T) d\nu  / \int_0^\infty B_\nu (T) d\nu.
 \end{equation}
  Here, the dust extinction coefficient $\chi_\nu$ is defined as 
$ \kappa_\nu + (1- g_\nu) \sigma_\nu$ with an asymmetry parameter $g_\nu$ and 
the scattering coefficient $\sigma_\nu$ \citep{Ishimaru1978,Min2009}. 
It should be noted that $B_\nu(T)$ in both Equations \ref{eq:kp} and \ref{eq:chi} is replaced 
by $\partial B_\nu(T)/\partial T$ when the Bjorkman-Wood method \citep{Bjorkman2001} 
is used.
 
\begin{figure*}
\includegraphics[width=1 \textwidth]{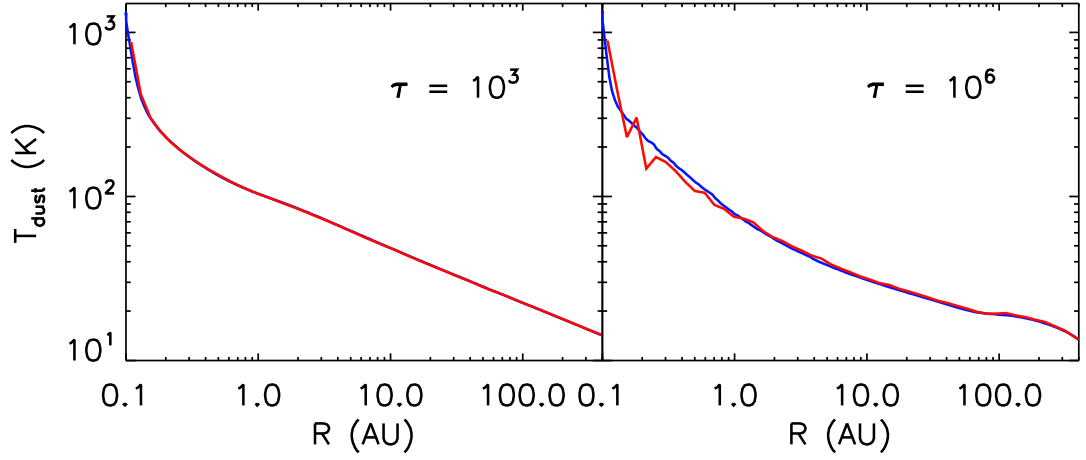}
\caption{Dust temperature distribution at the mid-plane in the optically thin 
case (left; $\tau = 10^3$) and in the optically thick case (right; $\tau=10^6$).
 The red and blue lines indicate the results from our model and  MCFOST 
 \citep{Pinte2006}, respectively.}
\label{fig:dust_benchmark}
\end{figure*}

%[Benchmark test] \\
We calculated the dust temperature for the protoplanetary disks in the 
benchmark test \citep{Pinte2009}\footnote{\url{https://ipag.osug.fr/~pintec/benchmark/}}.
Two cases were tested for isotropic scattering with disk masses of 
$3\times10^{-8}$ $M_\odot$ ($\tau=10^3$) and $3\times10^{-5}$ $M_\odot$ 
($\tau=10^6$).
Figure~\ref{fig:dust_benchmark} shows the dust temperature distribution at 
the midplane of the disk corresponding to our code (red) and for the MCFOST \citep[blue, ][]{Pinte2006}. 
In both cases, the agreement is very good and our results are within the 
deviations of other codes participating in the benchmark test.

%}}}1

\subsection{X-ray ionization rate}
X-ray photons are an important source of ionization at the intermediate column 
densities \citep{Igea1999,Offner2019}. In this work, we adopted the analytic 
formula given in \citet{Igea1999}.
The ionization rate can be expressed as a function of the column density 
of hydrogen nuclei vertically above ($N_{\rm a}$) and below ($N_{\rm b}$) 
at a given location in the disk \citep{Igea1999,Bai2009}:
\begin{eqnarray}
\zeta_X &=& L_{X,29} \left( \frac{r}{1\,{\rm AU}} \right)^{-2.2} \times \nonumber \\
&& ( \zeta_1 [ e^{-(N_{\rm a}/N_{\rm 1})^{\alpha_1}} +  e^{-(N_{\rm b}/N_{\rm 1})^{\alpha_1}}] + \nonumber \\
& & \zeta_2 [ e^{-(N_{\rm a}/N_{\rm 2})^{\alpha_2}} +  e^{-(N_{\rm b}/N_{\rm 2})^{\alpha_2}}] ),
\end{eqnarray}
where $L_{X,29} \equiv L_X/10^{29}$ erg s$^{-1}$, $L_{\rm X}$~is the X-ray luminosity,  
$\zeta_1 = 6 \times 10^{-12}$\, s$^{-1}$, $\zeta_2 = 10^{-15}$\, s$^{-1}$, $\alpha_1 = 0.4$, 
$\alpha_2 =0.65$, $N_1=1.5\times 10^{21}$\,cm$^{-2}$, and $N_2 = 7 \times 10^{23}$\,cm$^{-2}$. 

%newpage
\subsection{Chemistry} 
 %{{{1
Abundances of species were derived from the equations of chemical kinetics 
describing the formation and destruction of the species:

\begin{equation}
	\frac{dn_i}{dt} = \sum_{l,m}k_{lm}n_ln_m - n_i\sum_{l}k_{il}n_l +
	   \sum_{l}k_ln_l -n_i\sum_{l}k_l + k_i^{\rm des}n_i^{s} - k_i^{\rm ads}n_i
 \end{equation}
 \begin{equation}
	 \frac{dn_i^s}{dt} = \sum_{l,m}k^{s}_{lm}n_l^{s}n_m^{s} - n_i^{s}\sum_{l}k_{il}^{s}n_l^{s} 
	 - k_i^{\rm des}n_i^{s} + k_i^{\rm ads}n_i
\end{equation}
where $n_i$ and $n_i^{s}$ are the $i$-th species (cm$^{-3}$) in the gas and 
ice phases, respectively, 
$k_{lm}$ ($k_{il}$) and $k_i$ ($k_l$) are the gas-phase
reaction rates (in units of cm$^3$\,s$^{-1}$ and s$^{-1}$, respectively), 
$k_i^{\rm ad}$ and $k_i^{\rm des}$ denote the adsorption and desorption rates  
(s$^{-1}$), respectively, and $k_{lm}^s$ ($k_{il}^s$) is the surface reaction 
rate (cm$^3$\,s$^{-1}$). 
All reactions except the surface reaction are described in \citet{Lee2014}. 

 The chemical code of \citet{Lee2014} was developed from the Heidelberg 
``ALCHEMIC'' code \citep{Semenov2010}, which is based on the publicly available 
DVODPK (Differential Variable-coefficient Ordinary Differential equation solver
with the Preconditioned Krylov method GMRES for the solution of linear
systems) ODE package\footnote{\url{http://www.netlib.org/ode/vodpk.f}} and a 
high-performance sparse asymmetric MA28 solver from the Harwell Mathematical 
Software Library \footnote{\url{http://www.hsl.rl.ac.uk/}}. We replaced the 
MA28 solver with an updated solver version, MA48, which is faster than MA28 
by a factor of two.

We updated some reactions by considering the recent studies. 
When neutral species are adsorbed on the grain surface, we assume a sticking 
coefficient of 1.0, except for H and H$_2$. For H and H$_2$, we follow the gas-temperature-dependent formulae derived from laboratory experiments 
\citep{Matar2010,Chaabouni2012}.
The wavelength-dependent photodesorption experiments have shown that 
photodesorption is induced by  photoabsorption in the dominant ice component 
and energy transfer to the surface molecules
\citep{Fayolle2011,Fayolle2013,Bertin2012,Bertin2013}. Therefore, we used 
a single photodesorption rate of H$_2$O \citep[$Y_{\rm ph}= 10^{-3}$ molecule 
photon$^{-1}$,][]{Oberg2009} for all species. 

Changes in the dust opacity in the UV range due to the coagulation of dust grains 
was considered in the UV radiative transfer process. However, this also affects the UV photons
induced by cosmic rays \citep[][CRP]{Molano2012a,Molano2012b}. 
Although the opacities of some molecules can be comparable to the dust opacity 
\citep{Molano2012b}, we simply enhanced the photodissociation rates by CRP in the 
network using a scale factor of $\sigma_{\rm UV}/2\times10^{-21}$~cm$^{-2}$, 
where $\sigma_{\rm UV}$ is the UV extinction cross section for the dust grain 
in the model \citep{Molano2012a}. In addition, we also calculated the number of CRPs 
($N_{\rm CU}$) following the expression given by \citet{Molano2012a}:
\begin{equation}
\begin{split}
N_{\rm CU} = 12 500 \times \left(\frac{1}{1-\omega}\right) \left(\frac{\zeta_{\rm H_2}}{5\times 10^{-17}\,{\rm s}^{-1}}\right) \\
\times\left(\frac{2\times10^{-21}\, {\rm cm}^{-2}}{\sigma_{\rm UV}}\right)\left( \frac{n_{\rm H_2}/n_{\rm gas}}{0.5}\right),
\end{split}
\end{equation}
where $\omega$ is the grain albedo and $n_{\rm H_2}$ and $n_{\rm gas}$ are the 
number densities of molecular hydrogen and hydrogen nuclei, respectively. 
It should be noted here that this approximation is available when the gas opacity is much lower 
than the dust opacity \citep[see ][]{Molano2012b}.

 We also included the reactions with the vibrationally excited H$_2$ (v-H$_2$) in our model. 
Following \citet{Tielens1985} and  \citet{Bruderer2012}, the v-H$_2$ is a
vibrationally excited pseudo-level with an energy of 30163 K \citep{London1978}.
When two-body reactions with H$_2$ had an activation barrier, the exponential
factor $\gamma$ was replaced with max(0,$\gamma$- 30163 K) in the reactions with
v-H$_2$. The UV pumping rate was 8 times the H$_2$ photodissociation rate 
\citep{Sternberg2014}, and the spontaneous decay rate was 2$\times 10^{-7}$~s$^{-1}$ 
\citep{London1978}. 
The collision rates of H and H$_2$ were adopted from \citep{Tielens1985}.

For the grain surface chemistry of dust grains having a size of $a_\text{dust}$ 
and number density of $n_\text{dust}$, we considered only 
the Langmuir--Hinshelwood mechanism:
\begin{equation*}
\begin{split}
k_{lm}^s = &\kappa_{lm}\Bigg(k_\text{hop}^s(l) + k_\text{hop}^s(m)\Bigg) \\
&\times\frac{1}{N_\text{site} 4\pi a_\text{dust}^2 n_\text{dust}}~ [\text{cm}^3\text{s}^{-1}]
\end{split}
\end{equation*}
where $\kappa_{lm}$ is the probability that the reaction occurs. The thermal 
hopping rate are given by
\begin{equation}
	k_\text{hop}^s(l) = \sqrt{\frac{2N_\text{site} k_\text{B} E_\text{b}(l)}{\pi^2 m_l}} \, 
    \exp\Bigg(-\frac{E_\text{diff}(l)}{T_\text{dust}}\Bigg),
\end{equation}
where $N_\text{site}$ is the number density of the surface site 
($\simeq 1.5 \times 10^{15}\,\text{cm}^{-2}$),  $ k_\text{B}$ is the Boltzmann 
constant, $E_\text{b}(l)$ and $E_\text{diff}(l)$ are the binding and diffusion 
energies (in K) of the species $l$, respectively, and $m_l$ is its mass. 
We assumed that $E_\text{diff}(l) = 0.5 E_\text{b}(l)$ \citep{Garrod2011}.

When the reaction was exothermic and barrierless, we set $\kappa_{lm}=1$ 
\citep{Hasegawa1992}. 
For exothermic reactions with an activation barrier, 
denoted by $E_{\text{A},lm}$, we calculated $\kappa_{lm}$ as the result of the 
competition among reactions, hopping, and evaporation, as suggested by 
\citet{Garrod2011}:
\begin{equation*}
\kappa_{lm} = \frac{\nu_{lm} \kappa_{lm}^*}{\nu_{lm} \kappa_{lm}^* + k_\text{hop}^s(l) + k_\text{hop}^s(m)}.
\end{equation*}
where $\kappa_{lm}^*$ can be expressed as $\exp(-E_{\text{A},lm}/T_\text{dust})$ 
or the quantum mechanical probability for tunneling through a rectangular barrier 
of thickness $a$: $\kappa_{lm}^*=\exp[-2(a/ \hbar)(2\mu E_{\text{A},lm})^{1/2}]$, 
where $\mu$ is the reduced mass \citep[see][for details]{Hasegawa1992}. 
 Here, $\nu_{lm}$ was taken to be equal to the larger value of the characteristic 
frequencies of the two reactants $l$ and $m$ \citep{Garrod2013}.

%}}}1

\subsection{Self-Shielding and Line-Shielding} 

%FUV radiation{{{1
Far ultraviolet (FUV) radiation is emitted from the central star and the interstellar 
radiation field (ISRF). The FUV radiative transfer was calculated using 
a Monte Carlo method considering anisotropic scattering 
\citep[see ][]{Lee2014}, and the FUV fluxes from the two sources were treated 
separately. 
We used the opacities of the dust grains as given in Section~\ref{sec:dust}. 
The contributions of the UV flux along the radial, 
$G_{\rm dust}^{in}(ir,iz)$, and vertical direction, 
$G_{\rm dust}^{up}(ir,iz)$, were calculated for a given grid cell, $(ir,iz)$, in the disk.
Furthermore, a local column density, $N_{\rm local}(ir,iz)$, within a grid cell, 
$(ir,iz)$, was calculated as follows: 
\begin{equation}
\label{eq:lnc}
N_{\rm local} (ir,iz) = n_{\rm gas} (ir,iz) \frac{\sum_i I_i(s) \Delta s }{\sum_iI_i(s)}, 
\end{equation}
where $I_i(s)$ is the dust-attenuated intensity of a model photon $i$ 
penetrating the grid cell $(ir,iz)$, and $\Delta s$ is the path length traveled 
within the grid cell. 

%self-shielding effects
% {{{1
When taking the self-shielding effects of H$_2$, CO, C, and \mn into account,
their column densities along the path of UV photons should also be calculated.  
Thus, prior knowledge of their abundances in all grid cells is required. This problem could be solved using an iterative method 
\citep{vanZadelhoff2003,Bruderer2012}. 
In the 2D disk model, the UV photons from the central star mainly enter the 
grid cells through their inner as well as upper boundaries.
Therefore, we solved the problem in the consecutive order from the upper-innermost 
grid to the lower-outermost grid.

Then, the column densities of species X along the photon path from the central 
star to a given grid cell $(ir,iz)$ were derived as follows:
\begin{equation}
\label{eq:nc}
N_{\rm X}(ir,iz) = X_{\rm X}(ir,iz)\cdot N_{\rm local}(ir,iz)  + N_{\rm X}^{\rm ext} (ir,iz), 
\end{equation}
where the abundance of the species X is $X_{\rm X}(ir,iz)$, the local column 
density is $N_{\rm local}(ir,iz)$ (see Equation \ref{eq:lnc}), and the external 
column density of species X is 
 \begin{equation}
   N_{\rm X}^{\rm ext} (ir,iz) = \left\{\begin{array}{cl}
	   N_{\rm X}(ir-1,iz),               & \;if\,\,G_{\rm dust}^{in}(ir,iz) / G_{\rm dust}^{up}(ir,iz) > 10,  \\
    \frac{N_{\rm X}(ir-1,iz) \cdot G_{\rm dust}^{in}(ir,iz) + N_{\rm X}(ir,iz-1) 
	   \cdot G_{\rm dust}^{up}(ir,iz)}{G_{\rm dust}^{in}(ir,iz) + G_{\rm dust}^{up}(ir,iz) },   
	   & \; if\,\,  0.1  \leq G_{\rm dust}^{in}(ir,iz) / G_{\rm dust}^{up}(ir,iz) \leq 10,  \\
	   N_{\rm X}(ir,iz-1),               & \;if\,\,G_{\rm dust}^{in}(ir,iz) / G_{\rm dust}^{up}(ir,iz) < 0.1, 
  \end{array}\right.
 \end{equation}
 Here, we used the average value when the contributions of the vertical and radial 
directions to the dust-attenuated UV flux were comparable because the extreme 
cases could lead to incorrect values of the column density. For example, near the midplane in the 
outer disk, 
$G_{\rm dust}^{in}(ir,iz)$ is lower than $G_{\rm dust}^{up}(ir,iz)$, whereas $N_{\rm X}(ir-1,iz)$ is much 
higher than $N_{\rm X}(ir,iz-1)$, which leads to an unreasonably high value of the column density. 
 Furthermore, $N_{\rm X}(ir,iz+1)$ and $N_{\rm X}(ir-1,iz)$ are comparable, 
as shown in the bottom panels of Figure~\ref{fig:model2}, when 
$G_{\rm dust}^{up}(ir,iz)$ and $G_{\rm dust}^{in}(ir,iz)$ are similar. 
Therefore, the photodissociation rate of \mn with the self-shielding effect in 
our calculation is consistent with the sum of those in the vertical and radial 
directions even though the self-shielding function of \mn is nonlinear.

\begin{figure*}
\includegraphics[width=0.5\textwidth]{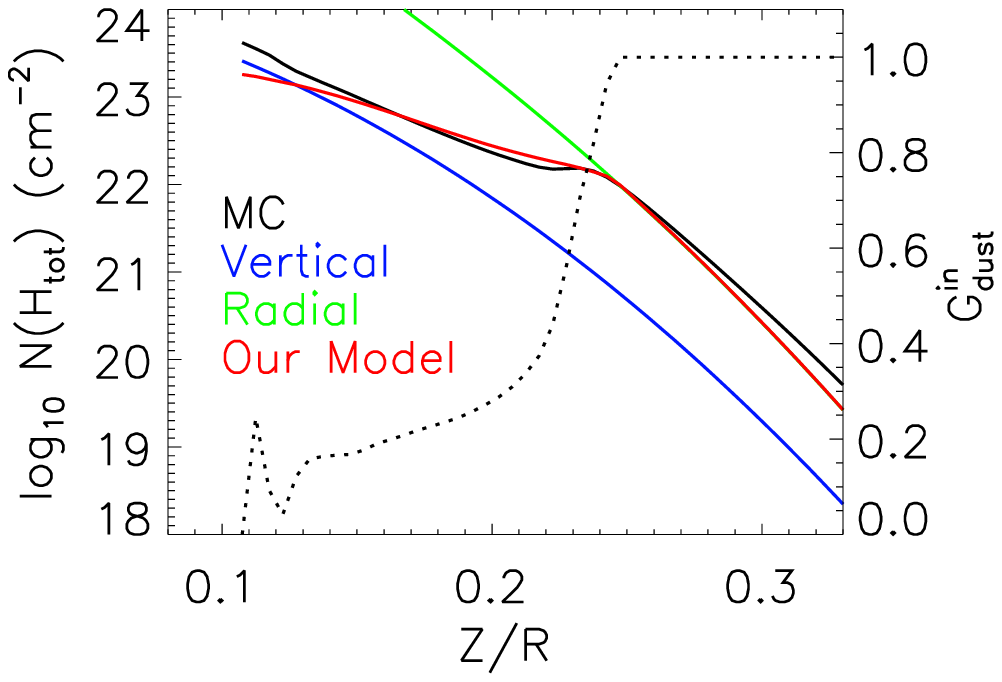}
\includegraphics[width=0.5\textwidth]{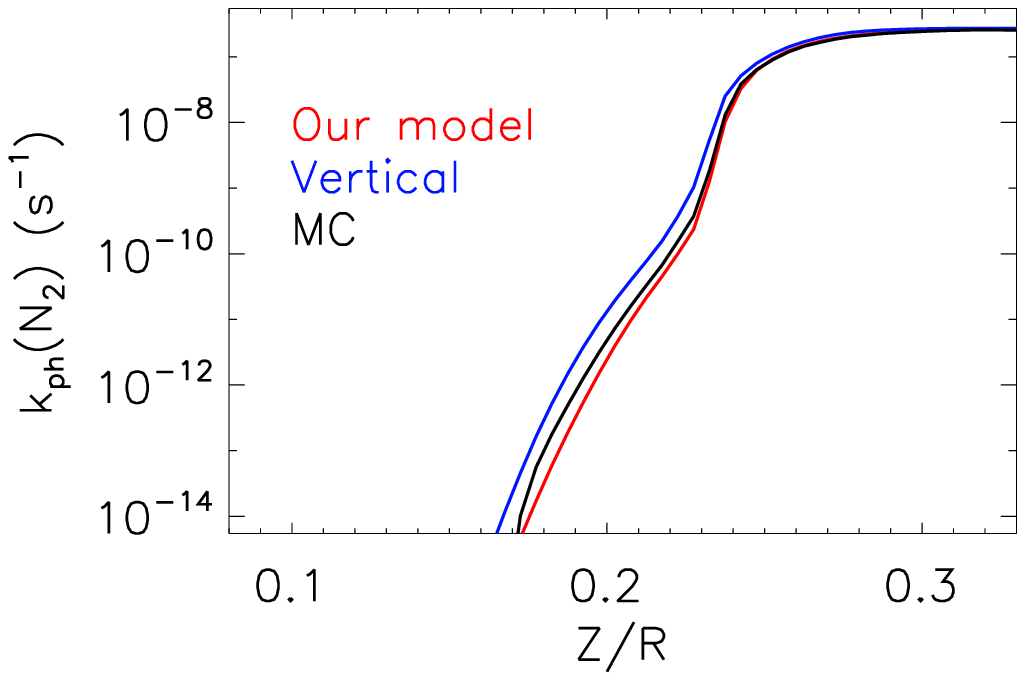}
\caption{ Total hydrogen column density from the central star ($N({\rm H_{tot}})$) 
along the vertical direction (left) and the photodissociation rate of \mn (right) 
at the radius of 40~AU in the ML model. Left: the black and red lines indicate the  
$N({\rm H_{tot}})$ calculated by the Monte Carlo method 
(Equation~\ref{eq:nc_full}) and with Equation~\ref{eq:nc}, respectively. 
 The column densities integrated radially from the central star and vertically 
from the upper atmosphere are presented in the green and blue lines, respectively. 
The black dotted line represents the contribution of UV photons along the radial direction 
to the UV flux for a given height ($G_{\rm dust}^{in}$). Right: the red and blue lines 
indicate the photodissociation rate of \mn with the self-shielding effect using the column
densities in the left panel with the same colors. The black line represents the 
photodissociation of \mn with the self-shielding effect, which is calculated along the path 
of model photons adopting the abundance distributions of atomic H, H$_2$, and \mn by 
the Monte Carlo method. 
}
\label{fig:check_nc}
\end{figure*}

 The left panel in Figure~\ref{fig:check_nc} shows the total hydrogen column 
densities from the central star to a given position along the vertical 
direction obtained using different methods at the radius of 40~AU. 
The black line ($N_{\rm MC}$) is the column density along the path of the photons 
from the central star to the grid $(ir,iz)$ obtained using the Monte Carlo simulation:
\begin{equation}
\label{eq:nc_full}
	{\rm ln}\, N_{\rm MC} (ir,iz) = \frac{\sum_i I_i(s) {\rm ln}\, N_i(s) }{\sum_i I_i(s)}, 
\end{equation}
 where $N_i(s)$ is the column density along the path of the model photon $i$ from 
the central star to the grid $(ir,iz)$. 
 In this method, ${\rm ln}\, N_i(s)$ is averaged instead of $N_i(s)$ because 
 the dynamical range of $N_i(s)$ is significantly large depending on the photon paths,
 owing to which it is difficult to obtain the averaged $N_i(s)$ numerically.
Our method using Equation~\ref{eq:nc} (the red line in the left panel in 
Figure~\ref{fig:check_nc}) derives the column density between the column 
densities integrated radially from the central star (green) and vertically 
from the upper atmosphere (blue) and is consistent with $N_{\rm MC}$ (black) 
within a factor of two. 
Therefore, our method was able to trace a realistic column density.

 The \mn photodissociation rates obtained using the different methods are shown in the right 
panel of Figure~\ref{fig:check_nc}. The red and blue lines indicate the \mn 
photodissociation rates obtained using the column densities shown with the same color in the left panel. The black line indicates the exact solution because the self-shielding 
effect is calculated along the path of the UV photons using the Monte Carlo method 
when the abundance distribution of atomic H, H$_2$, and \mn for the MLI3 model is 
adopted. Our method reproduces the same photodissociation rate in the 
atmosphere, whereas it slightly underestimates the photodissociation rate at 
a lower height where the \mn self-shielding effect is dominant as compared to the dust 
attenuation of UV photons.

The line-shielding effect for photodissociation by the stellar UV photons was also 
considered in our model. 
\citet{Heays2017} \footnote{\url{https://home.strw.leidenuniv.nl/~ewine/photo/}} 
provides tables for the line-shielding effect due to H, H$_2$, C, CO, and \mn. 
They assumed an excitation temperature of 100 K and a Gaussian Doppler 
broadening width of b = 3 km s$^{-1}$. The combined line-shielding functions were 
assumed to be a product of the line-shielding functions corresponding to the abovementioned 
five species. The column density was calculated using Equation~\ref{eq:nc}.
%}}}1

\subsection{Gas energetics}

%{{{1
In the disk, the properties of PAH and dust grains are different from those in the ISM.
For the photoelectric heating of large dust grains, we adopted the approximate 
formula for silicate grains in \citet{Kamp2001}, 
\begin{equation}
\Gamma_{\rm pe} = 2.5 \times 10^{-4}\,\sigma_{\rm UV}^{\rm tot}\,\epsilon\,G_{\rm dust}
\end{equation}
with
\begin{equation}
\epsilon = \frac{6\times 10^{-2}}{1 + 1.8 \times 10^{-3} x^{0.91}} 
  +\frac{y\left(10^{-4}\,T\right)^{1.2}}{1 + 10^{-2}x},
\end{equation}
 and 
 \begin{equation}
  y= \left\{\begin{array}{cl}
    0.7,    & \;if\,\, x\leq 10^{-4} \\
    0.36,   & \; if\,\,  10^{-4} \leq x \leq 1\\
    0.15,   &  \; if\,\, x > 1.
  \end{array}\right.
\end{equation}
Here, $G_{\rm dust}$ is the dust-attenuated UV strength in the Habing field, 
$\sigma_{\rm UV}^{\rm tot}$ is the total FUV absorption coefficient per unit volume,
and $x \equiv G_{\rm dust} \sqrt{T_{\rm gas}} / n_{\rm e}$ is the grain charge 
parameter, where $n_{\rm e}$ is the electron number density. It should be noted here that a new 
laboratory experiment has shown that the photoelectric yields of large dust 
grains are much larger than the theoretically expected values \citep{Abbas2006,Woitke2015}. 

The abundances of PAHs in the disk are much smaller (order of 1--2) than those 
in the ISM \citep{Geers2006}.
Thus, we scaled down the photoelectric heating rate by the PAH abundance ratio 
with respect to the standard PAH abundance in the ISM \citep{Tielens2008}, $f_{\rm PAH}$ and adopted its value to be 0.01.

The dust thermal accommodation was calculated, taking into account the
dust grain $i$ of size $a^i_{\rm dust}$, number density, $n^i_{\rm dust}$, and
dust temperature, $T^i_{\rm dust}$ \citep{Burke1983,Groenewegen1994,Woitke2015}, as follows:
\begin{eqnarray}
\Gamma_{\rm dust} - \Lambda_{\rm dust} &=& \sum_{i} 
	\pi (a^i_{\rm dust})^2 n^i_{\rm dust} \sqrt{\frac{8kT_{\rm gas}}{\pi m_{\rm H}}} n_{\rm gas} \alpha_{\rm acc} \nonumber \\
	& & \times \left(2kT^i_{\rm dust}- 2kT_{\rm gas} \right) \nonumber \\
	&\approx &  4 \times 10^{-12} n_{\rm gas} \sum_{i} \pi (a^i_{\rm dust})^2 n^i_{\rm dust}\alpha_{\rm acc} \nonumber \\
	&&\times\sqrt{T_{\rm gas}} (T^i_{\rm dust} - T_{\rm gas}),
\end{eqnarray}
with the efficiency for inelastic collision given by 
\begin{equation}
	\alpha_{\rm acc} \approx  0.1 + 0.35\, \exp \left(-\sqrt{\frac{T_{\rm gas} + T^i_{\rm dust}}{500 K}} \right).
\end{equation}
%}}}1

\newpage

\end{document}